\documentclass[a4paper,11pt]{article}
\usepackage[utf8]{inputenc}

\RequirePackage[OT1]{fontenc}
\RequirePackage{amsthm,amsmath,natbib}
\RequirePackage[colorlinks,citecolor=blue,urlcolor=blue]{hyperref}

\usepackage{graphicx,enumitem}

\usepackage{subcaption}

\usepackage{amsmath}
\usepackage[english]{babel}
\usepackage{amssymb}
\usepackage{amsfonts}
\usepackage{bm}
\usepackage{graphicx}
\usepackage{amsmath}
\usepackage{amsthm}
\usepackage{xcolor}
\usepackage{parskip}
\usepackage{multicol}
\usepackage{float}
\usepackage{rotating}
\usepackage{authblk}
\usepackage{geometry}
\newtheorem{teo}{Theorem}[section]

\newtheorem{lemma}[teo]{Lemma}
\newtheorem{prop}[teo]{Proposition}

\newmuskip\pFqmuskip
\newcommand*\pFq[6][8]{%
  \begingroup 
  \pFqmuskip=#1mu\relax
  \mathcode`\,=\string"8000
  \begingroup\lccode`\~=`\,
  \lowercase{\endgroup\let~}\pFqcomma
  {}_{#2}F_{#3}{\left[\genfrac..{0pt}{}{#4}{#5};#6\right]}%
  \endgroup
}
\newcommand{\pFqcomma}{\mskip\pFqmuskip}

\def\S{\mathbb{S}}
\def\R{\mathbb{R}}

\def\spacingset#1{\renewcommand{\baselinestretch}%
{#1}\small\normalsize} \spacingset{1}

\title{The $\mathcal{F}$-family of covariance functions: A Mat\'ern analogue for modeling random fields on spheres}

\author[1]{A. Alegr\'ia}
\author[1,5]{F. Cuevas-Pacheco}
\author[3]{P. Diggle}
\author[4]{E. Porcu}

\affil[1]{Departamento de Matem\'atica, Universidad T{\'e}cnica Federico Santa Mar{\'i}a, Valpara{\'i}so, Chile}
\affil[2]{Advanced Center for Electrical and Electronic Engineering, Universidad T\'ecnica Federico Santa Mar\'ia, Valpara\'iso, Chile}
\affil[3]{CHICAS, Lancaster Medical School, Lancaster University, United Kingdom}
\affil[4]{Department of Mathematics, Khalifa University of Science and Technology, Abu Dhabi, The Arab Emirates}
\affil[5]{Department of mathematical sciences, Aalborg University, Aalborg, Denmark}
\date{}

\begin{document}

\maketitle

\begin{abstract} 
The Mat{\'e}rn family of isotropic covariance functions has been central to the theoretical development and application of statistical models for geospatial data. For global data defined over the whole sphere representing planet Earth, the natural distance between any two locations is the great circle distance. In this setting, the Mat{\'e}rn family of covariance functions has a restriction on the smoothness parameter, making it an unappealing choice to model smooth data. Finding a suitable analogue for modelling data on the sphere is still an open problem. This paper proposes a new family of isotropic covariance functions for random fields defined over the sphere. The proposed family has  a parameter that indexes the mean square differentiability of the corresponding Gaussian field, and allows for any admissible range of fractal dimension. Our simulation study mimics the fixed domain asymptotic setting, which is the most natural regime for sampling on a closed and bounded set. As expected, our results support the analogous results (under the same asymptotic scheme) for planar processes that not all parameters can be estimated consistently. We apply the proposed model to a dataset of precipitable water content over a large portion of the Earth, and show that the model gives more precise predictions of the underlying process at unsampled locations than does the Mat{\'e}rn model using chordal distances.

\end{abstract}


\section{Introduction}
\label{sec:intro}

\subsection{Context}

The last decades have seen an unprecedented increase in the availability of georeferenced datasets of global extent, for example in the form of environmental monitoring networks or climate model ensembles \citep{castruccio1, PAF2016}. This increase in data-availability has in turn motivated the mathematical and statistical communities to develop models for random fields defined on the two-dimensional surface of the sphere, representing our planet. 

The Gaussian assumption implies that the finite dimensional distributions are completely specified through the mean and covariance function. Covariance functions are positive definite, and proving such a requirement involves substantial theoretical work. We refer the reader to \cite{schoenberg,gneiting2013,PBG16,berg-porcu} and \cite{WP} for the established theory about positive definite functions on the $d$-dimensional sphere of $\mathbb{R}^{d+1}$, for $d$ being a positive integer.  Also, comprehensive recent reviews can be found in \cite{jeong17} and in  \cite{PAF2016}. 

In spatial statistics, it is very common to assume the covariance function of a random field $Z$ to be isotropic, that is the covariance between $Z(\bm{x}_1)$ and $Z(\bm{x}_2)$ depends only on the distance between $\bm{x}_1$ and $\bm{x}_2$. 
For global data, the natural metric is the geodesic or great circle distance, defined as the length of the shortest arc joining two points located over the spherical shell. 

The Mat{\'e}rn covariance function is widely considered as the default choice for modelling spatial variation \citep{stein-book}. Its main attractive feature is its inclusion of a parameter that allows the user to control the fractal dimension and the mean square differentiability of the associated Gaussian process. In turn, this has been shown to be a fundamental aspect in evaluating predictive performance of covariance functions under infill asymptotics \citep{zhang}. The Mat{\'e}rn covariance has also a nice closed form expression for the associated spectral density, which is convenient for theoretical analysis of the properties of maximum likelihood (ML) estimators \citep{zhang}, approximate likelihood \citep{FGN,bevb:12, kaufman} and misspecified linear unbiased prediction \citep{stein-book} under infill asymptotics. A wealth of results is also available within SPDE's with Gaussian Markov approximations \citep{Lindgren} as well as in the numerical analysis literature. We refer the reader to \cite{SSS} for more details.

\subsection{The problem}

\cite{gneiting2013} shows that Mat{\'e}rn covariance functions are no longer positive definite on the sphere when coupled with the geodesic distance, unless a severe restriction is imposed on the smoothing parameter. Essentially, the Mat{\'e}rn covariance function can be used only for very rough realisations of the associated Gaussian process. Alternatively, the Mat{\'e}rn model can be adapted to the sphere by using the chordal distance, but such a choice would be suboptimal and we refer the reader to \cite{banerjee,gneiting2013,jeong-jun2,PBG16} for constructive criticism about the use of the chordal distance. 

The problem of obtaining a spherical analogue of the Mat{\'e}rn function that allows different degrees of differentiability has been addressed by using smoothing techniques of non-differentiable random fields \citep{jeong-jun} and by modelling the spectral representation of the covariance function on the sphere \citep{guinness}, providing the so-called circular Mat{\'e}rn model. The first approach does not allow for closed form expressions, whilst the second approach allows the use of linear combinations of hypergeometric polynomials to obtain closed form expressions, and of its series expansion for computations. Even though both approaches allow to index mean square differentiability, the lack of software makes the implementation difficult. Further, the computational problem increases when the smoothness of the covariance function at the origin decreases, as the convergence of the series becomes extremely slow. As a result, finding the analogue of the Mat{\'e}rn covariance function on the sphere is a challenging problem. In conclusion, the search for covariance functions on spheres that allow for a continuous parameterisation of smoothness is still elusive, and has been explicitly stated as an open problem in two collections of challenges posed by \cite{gneiting2013} and \cite{PAF2016}. Here we provide a solution to this problem.

\subsection{Our contribution}

We propose a class of covariance functions for $d$-dimensional spheres, that we term ${\cal F}$ class, having the same properties as the Mat{\'e}rn covariance function on planar surfaces. Specifically, the new class is specified through the Gauss hypergeometric function,  which has been widely studied in the numerical analysis literature \citep[see][and references therein]{Johansson2017arb}. For computation, we use the C library GSL \citep{galassi1996gnu}. However, libraries such as ARB \citep{Johansson2017arb} can be used as well.

We show that the new class has a parameter that allows for a continuous parameterisation of smoothness at the origin. Further, the same parameter allows for indexing fractal dimension of random surfaces on circles or spheres. Finally, we prove this class to admit several interesting closed form expressions that can be easily coded. Most applications deal with the two dimensional sphere, but for mathematical completeness we present our results over the $d$-dimensional sphere. 

The plan of the paper is the following. Section \ref{sec2} contains preliminaries needed for the subsequent presentation. Section \ref{sec3} introduces the ${\cal F}$-Family of covariance functions on spheres. We then study mean square differentiability and fractal dimension properties of Gaussian fields on spheres with the new class  of covariance functions. Section \ref{sec4} describes a simulation study to understand how well the parameters of the new covariance function can be estimated through ML. Our simulation study mimics the infill (or fixed domain) asymptotic framework \citep{stein-book} that is relevant for processes defined over closed and bounded sets. We especially focus on the estimation of scale, variance, and microergodic parameter \citep{stein-book} when the smoothing parameter is fixed. Our simulations suggest that consistent estimation of the scale and variance parameters is not achieveable, but that the microergodic parameter can be estimated consistently.  Section \ref{sec5} analyses a dataset corresponding to monthly averages of precipitable water content over a large portion of the planet Earth with a spatial resolution of $2.5$ degrees across longitudes and latitudes, available at \texttt{https://www.esrl.noaa.gov/psd/} \citep[See][]{kalnay1996ncep}. We show that the new class of covariance functions delivers better predictive performance on this dataset than both the Mat{\'e}rn covariance function based on chordal distance and the circular Mat{\'e}rn covariance function. The paper concludes with discussion. Technical details and generalisations that might be useful for future research are given in Appendices \ref{A.1} and \ref{A.2} respectively. The codes  associated with Sections \ref{sec4} and \ref{sec5}  are available at the following GitHub repository: \texttt{https:$//$github.com$/$FcoCuevas87$/{\text{F}}\_{}{\text{F}}$amil${\text{y}}\_{}{\text{c}}$ova}.

\section{Background} 
\label{sec2}

\subsection{Covariance functions and distances}
This section provides a background on random fields on the sphere, their covariance functions and their spectral representation.  For a positive integer  $d$,   $\mathbb{S}^d=\{\bm{x}\in\mathbb{R}^{d+1}, \|\bm{x}\|=1\}$ denotes the 
surface of the $d$-dimensional unit sphere embedded in $\R^{d+1}$, with $\|\cdot\|$ denoting Euclidean norm.  We shall sometimes refer to the Hilbert sphere $ \mathbb{S}^{\infty} = \{\bm{x} \in\mathbb{R}^{\mathbb{N}}, \|\bm{x}\|=1\}$. The natural metric on
$\mathbb{S}^d$ is the {\it great circle distance},
\begin{equation*}
\theta(\bm{x}_1,\bm{x}_2) = \arccos(\bm{x}_1^\top \bm{x}_2) \in [0,\pi],
\end{equation*}
 for $\bm{x}_1,\bm{x}_2\in\mathbb{S}^d$, where $\top$ denotes transpose.  The {\it chordal distance} on $\mathbb{S}^d$ is
\begin{equation}
\label{chordal} d_{{\rm CH}}(\bm{x}_1,\bm{x}_2) = \|\bm{x}_1-\bm{x}_2\| = 2 \sin \left (  \frac{\theta(\bm{x}_1,\bm{x}_2)}{2}\right ), \qquad \bm{x}_1,\bm{x}_2\in\mathbb{S}^d.
\end{equation}

We denote by $Z = \{Z(\bm{x}), \bm{x}\in\mathbb{S}^d\}$  a random field on $\mathbb{S}^d$, with constant mean
and covariance function  $C(\bm{x}_1,\bm{x}_2)={\rm cov}\{Z(\bm{x}_1),Z(\bm{x}_2)\}$, for $\bm{x}_1,\bm{x}_2\in\mathbb{S}^d$. The requirement for validity of a candidate function $C(\bm{x}_1,\bm{x}_2)$  to be a covariance function is that for any positive integer $n$, locations $\bm{x}_1,\hdots,\bm{x}_n \in \mathbb{S}^d$ and real numbers $c_1,\hdots,c_n$, 
\begin{equation}
\label{def_pos}
 {\rm Var}\left( \sum_{i=1}^n c_i Z(\bm{x}_i) \right)=\sum_{i,j=1}^n c_ic_j C(\bm{x}_i,\bm{x}_j) \geq 0.
 \end{equation}
Mappings $C$ that satisfy Equation (\ref{def_pos}) are called positive definite, or strictly positive definite if the inequality is strict for any non-zero collection of real numbers $c_1,\hdots,c_n$ and distinct locations $\bm{x}_1,\hdots,\bm{x}_n$.      
 
If in addition 
\begin{equation}
\label{isotropy} 
C(\bm{x}_1,\bm{x}_2) = \sigma^2 \psi(\theta(\bm{x}_1,\bm{x}_2)), \qquad \bm{x}_1,\bm{x}_2\in\mathbb{S}^d,
\end{equation}
for some value $\sigma^2>0$ and mapping $\psi:[0,\pi]\rightarrow \R$ such that $\psi(0)=1$,  then $C$ is called a geodesically isotropic covariance \citep{PAF2016}, and $\sigma^2$ is the variance of $Z$. Throughout, we use $\theta$ to denote great circle distance whenever no confusion can arise. Also, we shall not distinguish between positive and strict positive definiteness unless specifically required. We define $\Psi_d$ as the class of continuous functions $\psi$ associated with the covariance function $C$ on $\S^d$ through the identity (\ref{isotropy}). We also define $\Psi_\infty = \bigcap_{d= 1}^\infty \Psi_d$, with the strict inclusion relation 
\begin{equation*}
 \Psi_1 \supset \Psi_2 \supset \cdots \supset \Psi_d \supset \cdots \supset \Psi_{\infty}, 
\end{equation*} proved by \cite{gneiting2013}. 
\subsection{Spectral Theory}

Spectral representations for positive definite functions on spheres are equivalent to Bochner and Schoenberg's theorems in Euclidean spaces \citep[see][and references therein]{daley-porcu}.
\cite{schoenberg} showed that a continuous mapping $\psi:[0,\pi]\rightarrow \mathbb{R}$ belongs to the class $\Psi_d$ if and only if it can be uniquely written as
\begin{equation} \label{spectral_rep}
\psi(\theta) = \sum_{n=0}^{\infty} b_{n,d} \frac{P_{n}^{(d-1)/2}(\cos \theta)}{P_{n}^{(d-1)/2}(1)}, \qquad \theta \in [0,\pi], 
\end{equation}
where $P_{n}^\lambda$ denotes the $\lambda$-Gegenbauer polynomial of degree $n$ \citep[][22.2.3]{abramowitz1964handbook}, and $\{b_{n,d}\}_{n=0}^\infty$ is a probability mass function. 

\cite{schoenberg} also showed that
 $\psi$ belongs to the class $\Psi_\infty$ if and only if
\begin{equation}
\label{hilbert_sphere}
\psi(\theta) = \sum_{n=0}^{\infty} b_{n} (\cos \theta)^n, \qquad \theta \in [0,\pi],
\end{equation}
with $\{b_{n}\}_{n=0}^\infty$ being again a probability mass function. We follow \cite{daley-porcu} in calling the sequence $ \{ b_{n,d}\}_{n=0}^{\infty}$ in (\ref{spectral_rep}) a $d$-Schoenberg sequence, to emphasise the dependence on the index $d$ in the class $\Psi_d$. Analogously, we call
$\{ b_n \}_{n=0}^{\infty}$ a Schoenberg sequence. Fourier inversion allows for an explicit representation of the sequences $\{b_{n,d} \}_{n=0}^{\infty}$. Specifically, for $d = 1$ we have that \citep[see][]{gneiting2013}
\begin{equation} \label{d-schoenberg}
 b_{0,1} = \int_{0}^{\pi} \psi(\theta) {\rm d} \theta \qquad  \mbox{ and } \qquad  b_{n,1} = \frac{2}{\pi} \int_{0}^{\pi} \cos(n\theta) \psi(\theta) {\rm d} \theta, \mbox{ for }n \geq 1,
\end{equation}
whereas for $d \geq 2$ we have
\begin{equation} \label{d-schoenberg2}
 b_{n,d} = \kappa(n,d) \int_{0}^{\pi} \psi(\theta) P_{n}^{(d-1)/2}(\cos \theta)  \left (\sin \theta \right )^{d-1} {\rm d} \theta, 
\end{equation}
where $\kappa(n,d)$ is a positive constant \citep[see][]{gneiting2013}. 

\cite{lang-schwab} showed that the rate of decay of the $d$-Schoenberg sequence determines the regularity properties of the associated Gaussian field in terms of interpolation spaces and H{\"o}lder continuities of the sample paths. The $d$-Schoenberg sequences 
are useful in contexts as diverse as spatial statistics \citep{guinness}, equivalence of Gaussian measures and infill asymptotics  \citep{arafat}, approximation theory \citep{menegatto-peron, beatson, ziegel, massa}  and spatial point processes \citep{moller}. In practice, the $d$-Schoenberg sequence of a parametric model is rarely known and it must be computed via numerical integration. 

To build parametric models, it is common to use Equation \eqref{hilbert_sphere} and the relationship between probability mass functions and the associated probability generating functions. This procedure yields covariance functions models with known Schoenberg sequences. Table 1 in \cite{gneiting2013} provides a list of parametric models where the Schoenberg sequence is known \citep[see also][]{soub,PBG16}. 

The Negative Binomial probability distribution is defined through the coefficients
\begin{equation}
\label{peteee} 
b_{n}(\delta,\tau)= \binom{n+\tau-1}{n} \delta^n (1-\delta)^{\tau}, \qquad \delta \in (0,1), \tau >0, \quad n=0,1,\ldots.
\end{equation}
Using the coefficients $b_{n}(\delta,\tau)$ in concert with the expansion (\ref{hilbert_sphere}), \cite{gneiting2013} obtains the Negative Binomial family of members of the class $\Psi_{\infty}$, denoted 
 ${\cal N}_{\delta, \tau}$ throughout, and defined as 
 \begin{equation}
 \label{bin_family}
 {\cal N}_{\delta,\tau}(\theta) = \left ( \frac{1-\delta}{1-\delta \cos \theta } \right )^{\tau}, \qquad \theta \in [0,\pi].
 \end{equation}
One limitation of the Negative Binomial family is that its elements are infinitely differentiable at the origin, making them not very appealing for spatial interpolation \citep{stein-book}. 
Nevertheless, we shall use the function ${\cal N}_{\delta,\tau}$ as the starting point for the construction of our proposed family.

\subsection{The Mat{\'e}rn and the circular Mat{\'e}rn classes}

The Mat{\'e}rn class of covariance functions, ${\cal M}_{\nu, \alpha}$, is defined as \citep{stein-book}
\begin{equation}\label{matern}
 {\cal M}_{\nu, \alpha} (d_{{\rm CH}}) = \frac{2^{1-\nu}}{\Gamma(\nu)} \left ( \frac{d_{{\rm CH}}}{\alpha} \right )^{\nu} {\cal K}_{\nu}\left ( \frac{d_{{\rm CH}}}{\alpha} \right ), 
\end{equation}
where $d_{{\rm CH}}$ is the chordal distance as defined at ({\ref{chordal}), with
 $\alpha,\nu >0$ and ${\cal K}_{\nu}$ a modified Bessel function of the second kind of order $\nu$ \citep[][9.6.22]{abramowitz1964handbook}. 
The importance of the Mat{\'e}rn class stems from the parameter $\nu$ that controls the differentiability (in the mean square sense) of the associated Gaussian field. Specifically, for
any positive integer $k$,
 a Gaussian field with Mat{\'e}rn covariance function is $k$-times mean square differentiable if and only if $\nu > k$. Also,
 the Mat{\'e}rn function converges to the Gaussian kernel
as $\nu \rightarrow \infty$. When $\nu=k+ 1/2$, for $k$ a positive integer,  the Mat{\'e}rn covariance function simplifies into the product of an exponential covariance with a polynomial of order $k$. 
For instance, ${\cal M}_{1/2,1}(d_{{\rm CH}})= \exp(-d_{{\rm CH}})$ and ${\cal M}_{3/2,1}(d_{{\rm CH}})= \exp(-d_{{\rm CH}}) (1+d_{{\rm CH}})$. Observe that ${\cal M}_{\nu,\alpha}(0)=1$, so that ${\cal M}_{\nu,\alpha}$ is actually a correlation function, and must be premultiplied by a variance parameter to obtain a covariance function.

The Mat{\'e}rn covariance function is not in general a valid covariance function on ${\S}^2$. Lemma 1 in \cite{gneiting2013} shows that the function $\theta \mapsto {\cal M}_{\nu,\alpha}(\theta)$, $\theta \in [0,\pi]$, is not a member of the class $\Psi_1$ if $\nu > 1/2$. Thus, the Mat{\'e}rn class cannot be used to index arbitrary degrees of differentiability on spheres if coupled with the great circle distance. 

 \cite{guinness} have proposed the circular Mat\'ern covariance function, ${\cal C}_{\nu,\alpha}$, given by 
\begin{equation}\label{guinnes}
{\cal C}_{\nu,\alpha}(\theta) =  \frac{1}{S(\alpha,\nu)}\sum_{n=0}^\infty \frac{1}{(n^2+\alpha^2)^{\nu+1/2}}  \cos(n \theta), \quad \theta \in [0,\pi],
\end{equation}
\noindent with $S(\alpha,\nu) = \sum_{n=0}^\infty (n^2+\alpha^2)^{-(\nu+1/2)}$. 

Arguments in \cite{gneiting2013} show that ${\cal C}_{\nu,\alpha}$ belongs to the class $\Phi_3$. This model is an adaptation of the classical spectral representation of the Mat\'ern covariance on Euclidean spaces to the spherical case. \cite{guinness} show that the parameter $\nu$ controls the mean square differentiability of the associated Gaussian field on $\mathbb{S}^2$, and provide closed form expressions when $\nu$ is a half integer.

\section{The ${\cal F}$-family of covariance functions} \label{sec3}

We now let 
\begin{equation*}
{}_{2}F_{1}(a,b,c;z) = \sum_{n=0}^{\infty} \frac{(a)_{n}(b)_{n}}{(c)_{n}} \frac{z^{n}}{n!}, \quad |z| < 1, 
\end{equation*}
 \noindent denote the Gauss Hypergeometric function, where $(\cdot)_n$ denotes the Pochhammer symbol \citep[][6.1.22]{abramowitz1964handbook} and complex numbers $a$, $b$, $c$. As detailed in \citet[][p. 282]{whittaker1996course},  for $c > 0$, the Gauss hypergeometric function converges for all $|z| < 1$, and converges absolutely for $|z| = 1$ provided
\begin{equation}\label{condition_1}
\mathrm{Re}(c-a-b) > 0,
\end{equation}
\noindent where $\mathrm{Re}(x)$ denotes the real part of a complex number $x$. We now define the ${\cal F} = {\cal F}_{\tau,\alpha,\nu}$ family of functions through the identity
\begin{equation}
\label{cov_2f1}
\mathcal{F}_{\tau,\alpha,\nu}(\theta) = \frac{B(\alpha,\nu+\tau)}{B(\alpha,\nu)} {}_{2}F_{1}(\tau,\alpha,\alpha+\nu+\tau;\cos \theta), \qquad \theta \in [0,\pi],
\end{equation}
where $B(\cdot,\cdot)$ denotes the Beta function \citep[][6.2.1]{abramowitz1964handbook} and the parameters $\tau,\alpha$ and $\nu$ are strictly positive. We are now ready to illustrate the main result of this section. 
\begin{teo} \label{teo1}
Let $\tau$, $\alpha$ and $\nu$ be strictly positive. Then, the function ${\cal F}_{\tau,\alpha,\nu}$ defined through Equation (\ref{cov_2f1}) is a member of the class $\Psi_\infty$. Additionally, it can be written as a mixture of Negative Binomial covariance functions.
\end{teo}

{\bf Proof}. We give a constructive proof based on the following criterion that can be found in Lemma 1 in \cite{gneiting2013}, adapted to our notation. 
\begin{lemma} \label{Lemma1}
Let $q$ be a positive integer, $A \subseteq \mathbb{R}^{q}$ and let $\mu$ be a Borel probability measure on $A$. Let $\psi_{c}: [0,\pi] \to \R$ be an element of the class $\Psi_{\infty}$ for any $c \in A$. Then, the function $\psi:[0,\pi] \to \R$ defined by
\begin{equation}
\label{mixture}
\psi(\theta) = \int_{A}  \psi_{c}(\theta) \mu ({\rm d} c), \qquad \theta \in [0,\pi],
\end{equation} belongs to the class $\Psi_{\infty}$. 
\end{lemma}
We now consider the Negative Binomial family ${\cal N}_{\delta, \tau}$ as defined in Equation (\ref{bin_family}), and the Beta probability measure 
\begin{equation} \label{beta}
\mu_{\alpha,\nu}({\rm d} \delta)= \frac{1}{B(\alpha,\nu)} \delta^{\alpha-1} (1-\delta)^{\nu-1} {\rm d} \delta,  \qquad \delta \in (0,1), \quad \alpha,\nu >0.
\end{equation}
We invoke Lemma \ref{Lemma1} to claim that $\mathcal{F}_{\tau ,\alpha,\nu} \in \Psi_{\infty}$ because
\begin{equation}
\label{begin_proof}
\mathcal{F}_{\tau ,\alpha,\nu}(\theta) = \int_{(0,1)} {\cal N}_{\delta,\tau}(\theta) \mu_{\alpha,\nu} ({\rm d} \delta), \qquad \theta \in [0,\pi].
\end{equation}  In fact, direct inspection shows that, for $\theta \in [0,\pi]$, 
\begin{align} \label{ecuacion}
&  \frac{1}{B(\alpha,\nu)}\int_{0}^{1} \left(\frac{1-\delta}{1-\delta \cos \theta}\right)^{\tau} \delta^{\alpha-1}(1-\delta)^{\nu-1} \mathrm{d}\delta \nonumber \\
=& \frac{1}{B(\alpha,\nu)}\int_{0}^{1} \left(\sum_{n=0}^{\infty} \binom{n+\tau-1}{n}  (1-\delta)^{\tau}\delta^{n} (\cos \theta)^{n} \right) \delta^{\alpha-1}(1-\delta)^{\nu-1} \mathrm{d}\delta \nonumber  \\
=& \frac{B(\alpha,\nu+\tau)}{B(\alpha,\nu)}\sum_{n=0}^{\infty} \left( \int_{0}^{1}\delta^{n}\frac{\delta^{\alpha-1}(1-\delta)^{\tau + \nu - 1}}{B(\alpha,\nu+\tau)}\mathrm{d}\delta\right) \frac{(\tau)_{n}}{n!} (\cos \theta)^{n}, 
\end{align} where the second equality comes from (\ref{peteee}), and the last equality comes from dominated convergence. 
Note that the integral of the last expression corresponds to the $n$-th moment of a Beta distribution with shape parameters $\alpha$ and $\tau+\nu$, which is given by $(\alpha)_n/(\alpha+\tau+\nu)_n$ \citep{johnson1995continuous}. Thus,  it follows that
\begin{eqnarray}
\sum_{n=0}^{\infty} \frac{(\alpha)_{n}(\tau)_{n}}{(\alpha+\nu + \tau)_{n}} \frac{(\cos \theta)^{n}}{n!} &=& {}_{2}F_{1}(\tau,\alpha,\alpha+\nu+\tau;\cos \theta), \qquad \theta \in [0,\pi], 
\end{eqnarray}
which shows that (\ref{begin_proof}) and (\ref{cov_2f1}) agree as asserted. The proof is completed by invoking \cite{schoenberg}} theorem for the class $\Psi_{\infty}$, which is in turn described through our 
Equation (\ref{hilbert_sphere}).  \hfill $\Box$ \\

Notice that condition \eqref{condition_1} is achieved since $\nu > 0$. Also, ${\cal N}_{\delta,\tau}(0) = 1$ implies that $\mathcal{F}_{\tau,\alpha,\nu}(0) = 1$. This reproduces the result for $z = 1$ \citep[][equation 15.4.20]{olver2010nist}:
\begin{equation}\label{2f1in0}
{}_{2}F_{1}(\tau,\alpha,\alpha+\nu+\tau;1) = \frac{B(\alpha,\nu)}{B(\alpha,\nu+\tau)} = \frac{\Gamma(\nu)\Gamma(\alpha + \tau + \nu)}{\Gamma(\alpha + \nu)\Gamma(\tau + \nu)},
\end{equation}
provided that $\alpha, \tau$ and $\nu$ are positive real numbers. 

The proof of Theorem \ref{teo1} shows that the $\mathcal{F}$ class is obtained from the probability generating function of the so called Beta-Negative Binomial distribution \citep{johnson2005univariate}. The Schoenberg coefficients associated with the ${\cal F}$ class are uniquely defined as 
\begin{equation}
\label{schoenberg_Coeff}
b_n(\tau, \alpha, \nu) = \frac{B(\alpha,\nu+\tau)}{B(\alpha,\nu)}\frac{(\alpha)_{n}(\tau)_{n}}{(\alpha+\nu + \tau)_{n} n!}, \qquad n=0,1,\ldots.
\end{equation}
\noindent The associated $d$-Schoenberg coefficients $b_{n,d}(\tau,\alpha,\nu)$ are computed using Theorem 4.2(b) from \cite{moller} and are given in Appendix \ref{A.1}. Moreover, as a consequence of Stirling's formula, we derive
\begin{equation}\label{stirling}
b_n(\tau, \alpha, \nu) \sim M(\alpha,\tau,\nu) n^{-(1+\nu)}, \qquad n \to \infty,
\end{equation} 
\noindent where $M(\alpha,\tau,\nu)$ is a positive constant that depends on $\alpha,\tau$ and $\nu$. Here, for two functions $f,g:\mathbb{N} \to \R$, $f(n) \sim g(n)$ as $n \to \infty$ if and only if $\lim_{n \to \infty} f(n)/g(n) =1$. 

Equation \eqref{stirling} shows how the parameter $\nu$ controls the rate of decay of the Schoenberg coefficients. This is connected with the mean square differentiability \citep{lang-schwab} and the fractal dimension of the associated Gaussian random field \citep{hansen2015}.

\subsection{Mean square differentiability}

Defining mean square differentiability for a stochastic process defined on $\S^d$ needs care, as derivatives are not taken along straight lines (as in the case of processes defined over Euclidean spaces). \cite{guinness} provide a definition based on great circles, whilst \cite{lin2019extrinsic} define mean square differentiability through directional derivatives on manifolds. Albeit working on slightly different frameworks, the authors reach the same conclusion: for a Gaussian random field $Z$ that is isotropic on $\S^d$, with covariance function $C$ defined through Equation (\ref{isotropy}), for some member $\psi$ within the class $\Psi_d$, mean square differentiability of order $n$ of $Z$ is equivalent to the fact that the $2n$-th derivative of the even extensions of $\psi$ evaluated at zero, denoted $\psi^{(2n)}(0)$ throughout, exists and is finite.

Note that all the parametric families within the class $\Psi_{\infty}$ listed in Table 1 of \cite{gneiting2013}  are either nondifferentiable \citep[{\em e.g.}, the Sine Power family, see also][]{soub} or infinitely differentiable at the origin ({\em e.g.}, the Poisson and Negative Binomial families). Hence, the associated random fields are infinitely differentiable or nondifferentiable in the mean square sense.

The following result shows that the parameter $\nu$ controls the smoothness of a random field  with covariance $\mathcal{F}_{\tau,\alpha,\nu}(\theta)$. In what follows, $\lfloor x \rfloor$ denotes the largest integer less than or equal to $x \in \R$. 

\begin{prop} \label{prop1}
Let $d$ and $n$ be positive integers. Let $Z$ be an isotropic Gaussian random field on $\mathbb{S}^d$ with covariance function given by $\mathcal{F}_{\tau,\alpha,\nu}(\theta)$ as in Equation (\ref{cov_2f1}). Then, $Z$ is $n$ times mean square differentiable if and only if $\lfloor\nu\rfloor > n$.  
\end{prop}

\noindent {\bf Proof}. Our proof is based on direct application of Theorem 1 in \cite{guinness}: let $Z$ be a Gaussian random field on $\S^d$ with isotropic covariance function $C$ defined through the family $\mathcal{F}_{\tau,\alpha,\nu}$ defined at (\ref{cov_2f1}). Then, $Z$ is $n$ times mean square differentiable if and only if $\mathcal{F}^{(2n)}_{\tau,\alpha,\nu}(0)$ ({intended as even extension}) exists and is finite. To simplify the calculations below, we write 
$$f_{\delta,\tau}(\cos \theta) = (1-\delta)^{-\tau} {\cal N}_{\delta,\tau}(\theta) = (1-\delta \cos\theta)^{-\tau},$$
with ${\cal N}_{\delta,\tau}$ being the Negative Binomial family in Equation (\ref{bin_family}). Using this change of notation in concert with  the scale mixture representation (\ref{begin_proof}), we write $ \mathcal{F}_{\tau,\alpha,\nu}(\theta)$ as
\begin{equation}
\label{begin_proof2}
 \mathcal{F}_{\tau,\alpha,\nu}(\theta) = \frac{1}{B(\alpha,\nu)} \int_0^1 f_{\delta,\tau}(\cos\theta) \delta^{\alpha-1}(1-\delta)^{\tau+\nu-1} \text{d}\delta.
 \end{equation}
Let $n$ be a positive integer. To get the $2n$-th order derivative of $\mathcal{F}_{\tau,\alpha,\nu}$, we evoke dominated convergence to swap integrals with derivatives, so that we can write
\begin{equation} 
\label{que maravilla}
 \mathcal{F}^{(2n)}_{\tau,\alpha,\nu}(\theta) = \frac{1}{B(\alpha,\nu)} \int_0^1   \frac{\partial^{2n}  \{f_{\delta,\tau}(\cos\theta)\} }{\partial\theta^{2n}   } \delta^{\alpha-1}(1-\delta)^{\tau+\nu-1} \text{d}\delta.
\end{equation} 
Using Faa Di Bruno's formula, we find that
\begin{multline}
\label{der2n}
\mathcal{F}^{(2n)}_{\tau,\alpha,\nu}(\theta)  = \frac{1}{B(\alpha,\nu)} \int_0^1          \left( \sum g_{2n}(m_1,\hdots,m_{2n})     f_{\delta,\tau}^{(m_1+\cdots+m_{2n})}(\cos \theta)     \prod_{j=1}^{2n} (\cos^{(j)}\theta)^{m_j}  \right)   \\   \times \delta^{\alpha-1}(1-\delta)^{\tau+\nu-1} \text{d}\delta,
\end{multline}
where the sum is over all the $2n$-tuples of nonnegative integers $m_1,\ldots,m_{2n}$ satisfying the constraint 
\begin{equation}
\label{constraint}
1 \cdot m_1+ 2\cdot m_2 + \cdots + 2n\cdot m_{2n} = 2n.
\end{equation} 
Here, $g_{2n}(m_1,\hdots,m_{2n}) = (2n)! (m_1! 1!^{m_1} \cdots m_{2n}! (2n)!^{m_{2n}})^{-1}$ is a constant, and $\cos^{(j)}$ denotes the $j$-th derivative of the cosine function, {\em i.e.}, $\cos^{(j)}(\theta)$ is proportional to $\sin\theta$ if $j$ is odd, and to $\cos\theta$ if $j$ is even. 

When evaluating (\ref{der2n}) at $\theta=0$, the only terms that do not vanish in the sum are those with $m_1=m_3=\cdots=m_{2n-1} = 0$, since under this choice the sine functions are not involved in (\ref{der2n}). Hence, in Equation (\ref{der2n}) we only require computation of the derivatives of order $m_2+m_4+\cdots+m_{2n}$ of $f_{\delta,\tau}$. Also, the restriction (\ref{constraint}) simplifies to $m_2 + 2m_4 + \cdots + nm_{2n} = n$. We now observe that
$$  m_2+m_4+\cdots+m_{2n}  \leq m_2 + 2m_4 + \cdots + nm_{2n}  = n.$$
 Thus, last inequality {shows} that the maximum order of the derivatives of $f_{\delta,\tau}$ involved in (\ref{der2n}) is $n$ ({\em e.g.}, when $m_2=n$, and $m_4=\cdots = m_{2n} = 0$). 
 
 A straightforward calculation shows that the $j$-th derivative of the mapping $f_{\delta,\tau}(t)$, $t \in [-1,1]$, is given by $$ f_{\delta,\tau}^{(j)}(t) = (\tau)_{j+1}\delta^j (1-\delta t)^{-\tau-j}.$$
Therefore, evaluating (\ref{der2n}) at $\theta=0$, we obtain that ${\cal F}^{(2n)}_{\tau,\alpha,\nu}(0)$ exists and is finite provided finitely many Beta type integrals of the form
$$ \int_0^1        \delta^{\alpha + j -1}(1-\delta)^{\nu-j-1} \text{d}\delta, \qquad j\leq n,$$ are finite. 
{This is true provided} the parameters $\alpha+j$ and $\nu-j$ are positive, for all $j\leq n$, which is equivalent to $\nu > n$. The proof is completed. 
\hfill $\Box$ \\

To prove that $\cal F_{\tau,\alpha,\nu}$ allows for a continuous parameterisation of smoothness through the parameter $\nu$, we need to assess the limiting behaviour of ${\cal F}_{\tau,\alpha,\nu}$ when the parameter $\nu$ goes to infinity in some sense. This needs some care as discussed at the end of Section \ref{sec-para}. 

\subsection{Fractal dimensions}

As noted by \cite{hansen2015}, the roughness or smoothness of a surface at an infinitesimal scale is quantified by the Hausdorff or fractal dimension, $D$, which for a surface in $\mathbb{R}^3$ must lie within the interval $[2,3)$, attaining the lower limit when the surface is differentiable. Moreover, \cite{hansen2015} provide a method based on kernel smoothing to obtain random fields with the desired fractal index. However, such a procedure does not allow for closed form expressions for the resulting covariance function.

An isotropic random field $Z$ on the sphere $\S^2$ with correlation function $C$ defined through some member $\psi$ within the class $\psi \in \Psi_2$, has fractal index $a \in (0,2]$ if there exists a constant $b>0$ such that 
\begin{equation} \label{fractal_dim_def}
\lim_{\theta \searrow 0} \frac{\psi(0) - \psi(\theta)}{\theta^{a}} = b,
\end{equation} 
\noindent where $\lim_{\theta \searrow 0}$ denotes the limit taken from the right. The fractal index exists for most parametric families of correlation functions, in which case the fractal dimension $D$ and fractal index $a$ are related by $D=3-a/2$, so that $a=2$ and $a\to 0$ correspond to extreme smoothness and roughness, respectively.
\begin{prop} \label{fractal}
Let $a$ be the fractal index defined at (\ref{fractal_dim_def}). Let $Z$ be an isotropic Gaussian random field on $\S^2$, with covariance function given by $\mathcal{F}_{\alpha,\tau,\nu}(\theta)$ as in Equation (\ref{cov_2f1}). 
Then, the fractal index $a$ of $Z$ is
$$
a=
\left\{\begin{array}{lr}
        2 \nu, & \text{if } 0 < \nu < 1, \\
        2, & \text{if } \nu > 1, 
     \end{array}\right.
$$ and $a$ is not defined  when $\nu=1$. 
\end{prop}

Before providing a formal proof, some comments are in order. According to Proposition \ref{fractal}, a realisation of a Gaussian field with covariance function belonging to the ${\cal F}$-Family is smooth when $\nu > 1$, and rough when $\nu$ is smaller than one. In fact, $D=3-\min(2\nu,2)$ when $\nu < 1$, and $D=2$ whenever $\nu > 1$. The same result is obtained for the Mat{\'e}rn family of covariance functions for random fields defined on $\R^2$. The fact that we can characterise the fractal dimension through the parameter $\nu$ gives an additional way to interpret the effect of this parameter on the properties of the process $Z$. We now prove formally the assertion above.

\noindent {\bf Proof}.
We provide a proof by direct construction. We need to evaluate the limit at Equation (\ref{fractal_dim_def}) with $\psi(\theta)= \mathcal{F}_{\alpha,\tau,\nu}(\theta)$, $\theta \in [0,\pi]$. We apply  L'H$\hat{\text{o}}$pital's rule to obtain
\begin{equation} \label{limite1}
\lim_{\theta \searrow 0} \frac{\mathcal{F}_{\alpha,\tau,\nu}(0) - \mathcal{F}_{\alpha,\tau,\nu}(\theta)}{\theta^{a}} = \frac{B(\alpha,\tau+\nu)}{B(\alpha,\nu)}\lim_{\theta \searrow 0} \frac{\sin\theta \;  _{2}F_{1}(\alpha+1,\tau+1,\alpha+\tau+\nu+1,\cos \theta)}{a\theta^{a-1}}.
\end{equation} 
We now inspect \eqref{limite1} depending on $\nu$. Let us first assume that $\nu>1$. Then, by \eqref{2f1in0}, the limit (\ref{limite1}) exists for $a=2$. 
Next, for $\nu < 1$, we have 
\begin{eqnarray} \label{solange}
& & \lim_{\theta \searrow 0} \frac{\mathcal{F}_{\alpha,\tau,\nu}(0) - \mathcal{F}_{\alpha,\tau,\nu}(\theta)}{\theta^{a}} \nonumber \\ &=& \frac{B(\alpha,\tau+\nu)}{aB(\alpha,\nu)}\lim_{\theta \searrow 0} \frac{\sin\theta}{\theta} \; \frac{ _{2}F_{1}(\alpha+1,\tau+1,\alpha+\tau+\nu+1,\cos \theta)}{(1-\cos \theta)^{\nu-1}} \frac{(1-\cos \theta)^{\nu-1}}{\theta^{a-2}} \nonumber \\
&=& \frac{B(\alpha,\tau+\nu)}{aB(\alpha,\nu)}\frac{\Gamma(\alpha+\tau+\nu+1)\Gamma(1-\nu)}{\Gamma(\alpha+1)\Gamma(\tau+1)} \lim_{\theta \searrow 0} \left(\frac{1-\cos \theta}{\theta^{(a-2)/(\nu-1)}}\right)^{\nu-1},
\end{eqnarray}
where the last  equality is obtained as direct application of Equation 15.4.23 in \citet{olver2010nist}:
\begin{equation*}
\lim_{x\nearrow 1} \frac{{}_2 F_{1}(\alpha+1,\tau+1,\alpha+\tau+\nu+1,x)}{(1-x)^{\nu-1}} = \frac{\Gamma(\alpha+\tau+\nu+1)\Gamma(1-\nu)}{\Gamma(\alpha+1)\Gamma(\tau+1)}. \label{eq:limit_olver1mx}
\end{equation*}
 Also, notice that 
$$\lim_{\theta \searrow 0} \left(\frac{1-\cos \theta}{\theta^{(a-2)/(\nu-1)}}\right)^{\nu-1} = \left\{\begin{array}{cc}
\frac{1}{2^{\nu-1}}& \mbox{ if }a = 2\nu,\\
 & \\ 
0 &  \mbox{ if }a < 2\nu, \\
& \\
\infty &  \mbox{ if }a > 2\nu. \\
& 
\end{array}\right.$$
Finally, for $\nu = 1$, we have that 
\begin{eqnarray*}
& & \lim_{\theta \searrow 0} \frac{\mathcal{F}_{\alpha,\tau,1}(0) - \mathcal{F}_{\alpha,\tau,1}(\theta)}{\theta^{a}} \nonumber \\ &=& \frac{B(\alpha,\tau+1)}{aB(\alpha,1)}\lim_{\theta \searrow 0} \frac{\sin\theta}{\theta}\frac{ _{2}F_{1}(\alpha+1,\tau+1,\alpha+\tau+2,\cos \theta)}{-\log(1-\cos \theta)} \frac{[-\log(1-\cos \theta)]}{\theta^{a-2}}\\
&=& \frac{B(\alpha,\tau+1)}{aB(\alpha,1)}  \frac{\Gamma(\alpha+\tau+2)}{\Gamma(\alpha+1)\Gamma(\tau+1)} \lim_{\theta \searrow 0} \frac{-\log(1-\cos \theta)}{\theta^{a-2}},
\end{eqnarray*}
where the last equality is obtained by using the following limit \citep[][Equation 15.4.21]{olver2010nist}:
\begin{equation*}
\lim_{x\nearrow 1} \frac{{}_2 F_{1}(\alpha+1,\tau+1,\alpha+\tau+2,x)}{-\log(1-x)} = \frac{\Gamma(\alpha+\tau+2)}{\Gamma(\alpha+1)\Gamma(\tau+1)}.
\end{equation*}
Then, noticing that
\begin{equation}
\lim_{\theta \searrow 0} \frac{-\log(1-\cos \theta)}{\theta^{a-2}} = \left\{\begin{array}{cc}
\infty& \mbox{ if }a = 2,\\
0 &  \mbox{ if }0 < a < 2, \\ 
\end{array}\right.
\end{equation}
we conclude that the fractal index does not exists for $\nu = 1$. \hfill $\Box$

\subsection{Special cases}  \label{sec-para}

We now show that the $\mathcal{F}$-family admits closed form expressions when the smoothing parameter is of the form $\nu=1/2+k$, for positive integers $k$. We use the following recurrence formula \citep[][Equation 15.5.18]{olver2010nist}: 
\begin{eqnarray}
\label{recurrencia}
  _2F_1(a,b,c+1; z) &=& \frac{1}{(c-a)(c-b)z}\Bigg ( c(1-c)(z-1)   _2F_1(a,b,c-1; z)  - \nonumber \\
  &-& c(c-1-\{2c-a-b-1\}z)   _2F_1(a,b,c; z) \Bigg ),
\end{eqnarray}
for $0 < |z| < 1$. Equation (\ref{recurrencia}) is undefined at $z=0$ or $z=1$. However, by definition, the Gauss Hypergeometric function is identically equal to 1 at $z=0$ and, under condition \eqref{condition_1}, the left hand side of \eqref{recurrencia} is well defined at $z=1$. 

To iterate the recurrence in Equation (\ref{recurrencia}), we need to provide two initial conditions on the right hand side of the equation.  Let $g(z) = \frac{1}{2} + \frac{1}{2}(1-z)^{1/2}$ for $|z|\leq 1$. Also, for $\alpha>0$, let
$a=\alpha$ and $b=\alpha+1/2$. We then have the following  identities \citep[][Equation 15.4.17-18]{olver2010nist}:
\begin{eqnarray}
\label{caso_particular1}
  _2F_1(\alpha,\alpha + 1/2,2\alpha; z)   &  = & \frac{1}{(1-z)^{1/2}}g(z)^{1-2\alpha}, \\
  \label{caso_particular2}
 _2F_1(\alpha,\alpha + 1/2,2\alpha + 1; z)   &  = & g(z)^{-2\alpha}. 
\end{eqnarray}
In particular, Equation (\ref{caso_particular2}) provides a covariance function that is continuous but not differentiable at the origin. Thus, any Gaussian random field with such a covariance function would be mean square continuous but nondifferentiable. 

\begin{figure}
\centering
\includegraphics[scale=0.55]{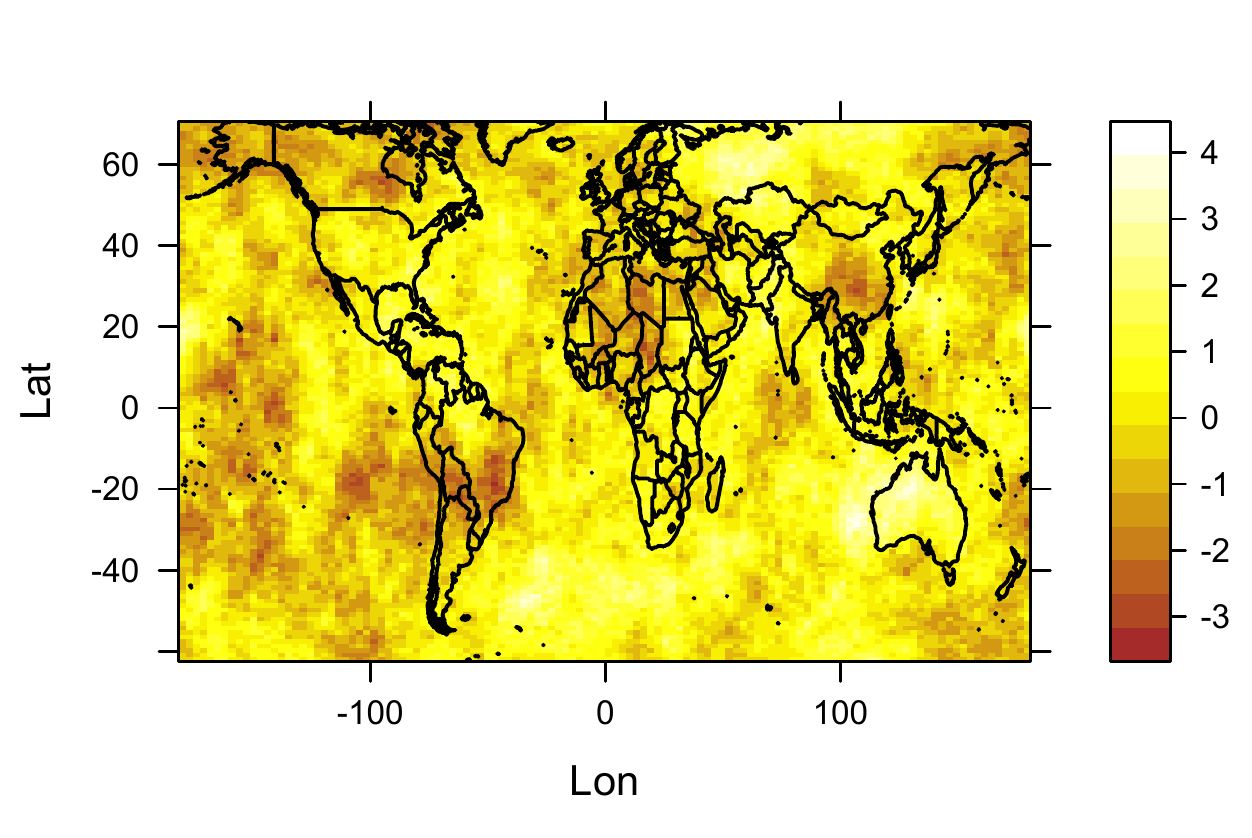}  \hspace{1cm}  \includegraphics[scale=0.55]{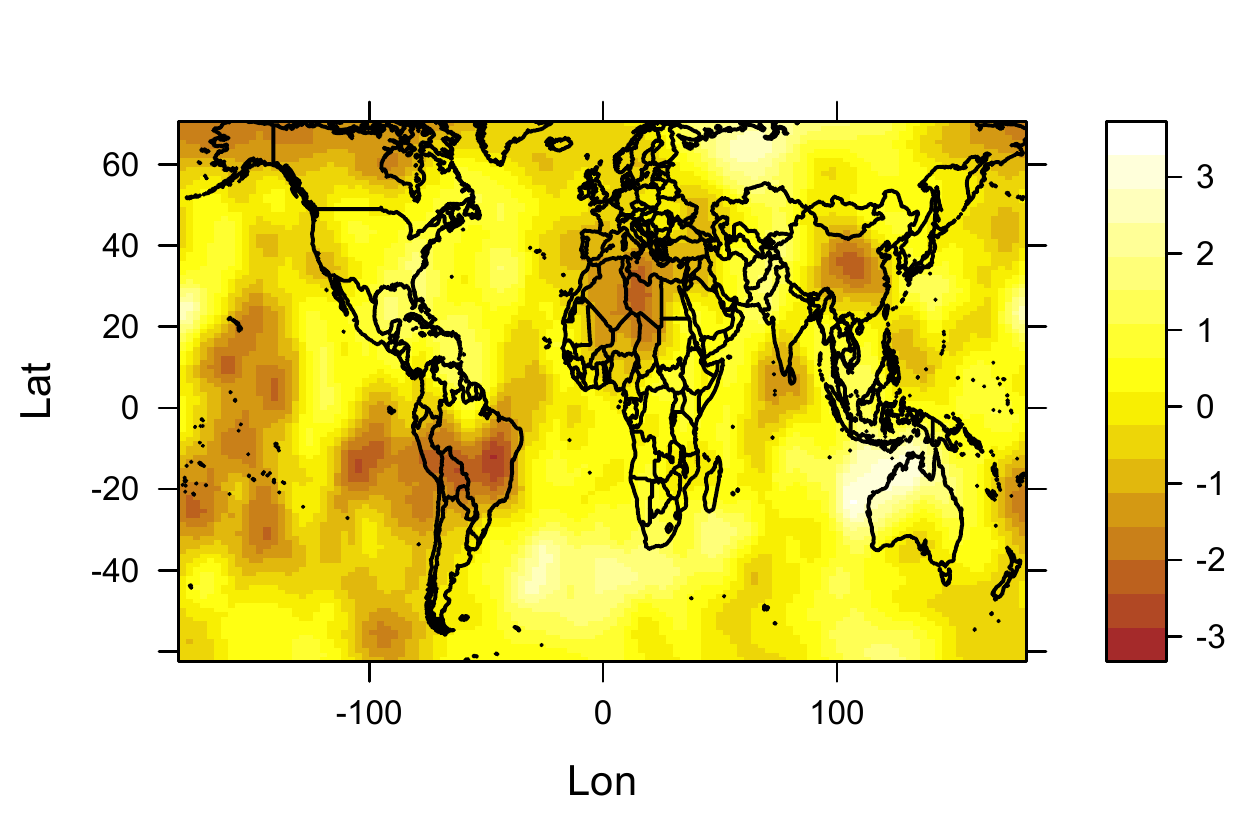}
\caption{Simulated datasets from the $\mathcal{F}$-family with $\sigma^2=1$ and an approximated practical range of $4675$ kilometers. We consider $\nu=1/2$ (Left) and $\nu=5/2$ (Right). We have used the same random seed for both realisations. }
\label{realizaciones}
\end{figure}

To obtain special cases with higher degrees of differentiability at the origin, we  can combine Equation (\ref{recurrencia}) with the special cases (\ref{caso_particular1}) and (\ref{caso_particular2}) yielding a once differentiable covariance function:
\begin{equation*}
  _2F_1(\alpha,\alpha + 1/2,2\alpha+2; z) = \frac{(2\alpha+1)g(z)^{-2\alpha} p(z)}{(\alpha+1)(\alpha+1/2)z},
\end{equation*}
where $p(z) = -(\alpha+1/2)(1-z) + \alpha(1-z)^{1/2} + 1/2$.  Iterating the formula once more, we obtain a covariance function generating twice mean square differentiable Gaussian random fields:
\begin{eqnarray}
\label{caso_particular3}
  _2F_1(\alpha,\alpha + 1/2,2\alpha + 3; z) &=& \frac{(2\alpha+2)g(z)^{-2\alpha}}{(\alpha+2)(\alpha+3/2)z^2} \Bigg ( \frac{(-2\alpha-1 + \{ 2\alpha+5/2\}z)(2\alpha+1)  p(z) }{(\alpha+1)(\alpha+1/2)}   \nonumber \\
  &+&    (2\alpha+1)(1-z)z    \Bigg ).
\end{eqnarray}
Figure \ref{realizaciones} depicts two realisations from the $\mathcal{F}$-family, with $\nu=1/2$ and $\nu=5/2$. We have reported the realisations over a planar grid of latitudes and longitudes in order to provide a better visualisation. To control the variance, we have used rescaled versions of Equations (\ref{caso_particular2}) and  (\ref{caso_particular3}). We choose $\alpha$ such that both covariance functions have an approximate practical range (the great circle distance at which the correlation is identically equal to $0.05$) of $4675$ kilometers.   We shall use this parameterisation in Sections \ref{sec5} and \ref{sec6}. 

\subsection{Limit cases} \label{limit}

Note how the proof of Theorem \ref{teo1} emphasises that ${\cal F}_{\tau, \alpha, \nu}$ is the scale mixture of ${\cal N}_{\delta, \tau}$ with a Beta distribution with parameters $\alpha$ and $\nu$, where the mixture is taken with respect to $\delta \in (0,1)$. Our next result illustrates the limiting behaviour of the ${\cal F}$ covariance when $\nu,\alpha \rightarrow \infty$, in such a way that $\nu/\alpha$ is asymptotically constant.
\begin{prop}\label{prop2}
Let ${\cal F}_{\tau,\alpha,\nu}$ be the family defined through (\ref{cov_2f1}). Let $\{ \alpha_n \}_{n=1}^{\infty}$ and $\{ \nu_n\}_{n=1}^{\infty}$ be two positive and increasing sequences such that $\nu_n/\alpha_n$ tends to a positive and finite constant, $\kappa$, when $n \to \infty$. Let $\mathcal{N}_{\delta,\tau}$ be the Negative Binomial family as defined in (\ref{bin_family}).
Then, for each $\theta\in[0,\pi]$, $$ \mathcal{F}_{\tau,\alpha_n,\nu_n}(\theta) {\longrightarrow}  \mathcal{N}_{1/(1+\kappa),\tau}(\theta),  $$ when $n \rightarrow \infty$. 
\end{prop}
{\bf Proof}.  Let $\{ \Delta_n \}_{n=1}^{\infty}$ be a sequence of random variables such that, for each $n \in \mathbb{N}$, $\Delta_n$ is distributed  according to $\mu_{\alpha_n,\nu_n}$ as defined through Equation (\ref{beta}). Hence, 
\begin{equation} \label{cuevas_se_casa}
{\rm E}({\Delta}_n) = \frac{1}{1 + \nu_n/\alpha_n} \qquad \text{and} \qquad	 {\rm Var}({\Delta_n}) = \frac{ \nu_n/\alpha_n }{(1+\nu_n/\alpha_n)^{2}(\alpha_n + \nu_n + 1)}.
\end{equation}
We invoke again the scale mixture argument in Equation (\ref{begin_proof}) to  write
\begin{equation*}
\label{mixture2}
\mathcal{F}_{\tau,\alpha_n,\nu_n}(\theta)    =    \int_{0}^{1} \mathcal{N}_{\delta,\tau}(\theta) \frac{\delta^{\alpha_n-1}(1-\delta)^{\nu_n-1}}{B(\alpha_n,\nu_n)} \mathrm{d}\delta, \qquad \theta \in [0,\pi].\   
\end{equation*}
Hence,
\begin{eqnarray}
 \lim_{n \to \infty} \mathcal{F}_{\tau,\alpha_n,\nu_n}(\theta) &=&   \lim_{n \to \infty}  \int_{0}^{1} \mathcal{N}_{\delta,\tau}(\theta) \frac{\delta^{\alpha_n-1}(1-\delta)^{\nu_n-1}}{B(\alpha_n,\nu_n)} \mathrm{d}\delta \nonumber \\
 &=& \int_{0}^{1} \mathcal{N}_{\delta,\tau}(\theta)  \lim_{n \to \infty}  \frac{\delta^{\alpha_n-1}(1-\delta)^{\nu_n-1}}{B(\alpha_n,\nu_n)} \mathrm{d}\delta, \label{Delta_measure}
\end{eqnarray}
where the last line is justified by dominated convergence. Using the fact that $\nu_n/\alpha_n$ tends to $\kappa$, as $n \to \infty$, we have that ${\rm E}({\Delta}_n)$ in Equation (\ref{cuevas_se_casa}) tends to $1/(1+\kappa)$, whereas ${\rm Var}(\Delta_n)$ tends to zero. This implies that the sequence of random variables $\{ \Delta_n \}_{n=1}^{\infty}$ converges, {in the mean square sense,} to a random variable having a probability mass with a single atom at $1/(1+\kappa)$. This implies that the right hand side of (\ref{Delta_measure}) is identically equal to ${\cal N}_{1/(1+\kappa),\tau}(\theta)$.
\hfill $\Box$

\section{Simulation study} \label{sec4}

The ML method is generally considered best for estimating the parameters of statistical models, although the theoretical justification for this stems primarily from the asymptotic properties of ML estimators. In the present context,
the study of asymptotic properties of ML estimators is complicated by the fact that the only physically sensible asymptotic regime for a process on the unit sphere is fixed domain asymptotics, {\em i.e.}, increasingly dense sampling of $Z$ on its fixed domain, the unit sphere.

It is generally the case that for spatially continuous processes, under fixed domain asymptotics,
prediction is consistent but parameter estimation is not \citep{stein-book}.
The rationale for this simulation study is therefore to explore the finite sample behaviour of ML estimates for the $\mathcal{F}$-Family of covariance functions. 

A key theoretical result for fixed domain asymptotics is the equivalence of Gaussian measures associated with random fields defined over bounded sets of $\mathbb{R}^d$  \citep{SY73}. Equivalence of Gaussian measures has specific consequences for both ML estimation and for kriging predictions. First, equivalence implies that the ML estimates of the parameters of a given class of covariance functions cannot be estimated consistently. 
Second, the misspecified kriging  predictor under the wrong covariance model is asymptotically equivalent to the kriging predictor
 under the true covariance. For the Mat\'ern covariance function, using Euclidean distance and assuming the smoothing parameter to be fixed, \cite{zhang} shows that the scale and the variance cannot be estimated consistently. Instead, a specific function of the variance and the 
scale (called microergodic parameter -- see below) can be estimated consistently.

\subsection{Maximum likelihood estimates}

We first study the influence of the correlation range and  differentiability on the variability of the ML estimates. We parameterise
the $\mathcal{F}$-Family of covariance functions as
\begin{equation} \label{sergio-ramos}  {\cal F}_{1/\alpha, 1/\alpha+0.5, \nu}(\theta) =      \sigma^2 \frac{ \Gamma(\frac{1}{\alpha}+\frac{1}{2}+\nu)\Gamma(\frac{1}{\alpha}+\nu)   }{ \Gamma(\frac{2}{\alpha}+\frac{1}{2}+\nu)\Gamma(\nu)    }           {}_{2}F_{1}\left(\frac{1}{\alpha}, \frac{1}{\alpha} + \frac{1}{2}, \frac{2}{\alpha} + \frac{1}{2}+ \nu, \cos\theta\right), \end{equation}
with $0\leq \theta \leq \pi$. An increase in $\alpha$ corresponds to increasing the correlation range. We set $\sigma^{2} = 1$ and
consider four scenarios for $\alpha$ and $\nu$: (a)  $(\alpha,\nu) = (0.3,0.5)$;  (b) $(\alpha,\nu) = (0.6,0.5)$;   (c) $(\alpha,\nu) = (0.3,2.5)$; (d) $(\alpha,\nu) = (0.6,2.5)$. Scenarios (a) and (b)  correspond to a continuous, non-differentiable random field, whereas Scenarios (c) and (d) to a twice differentiable random field.  Each simulated realisation generates $N=256$ data-values on a $14\times 14$  grid of longitudes and latitudes.

\begin{figure}
\centering
\includegraphics[scale=0.35]{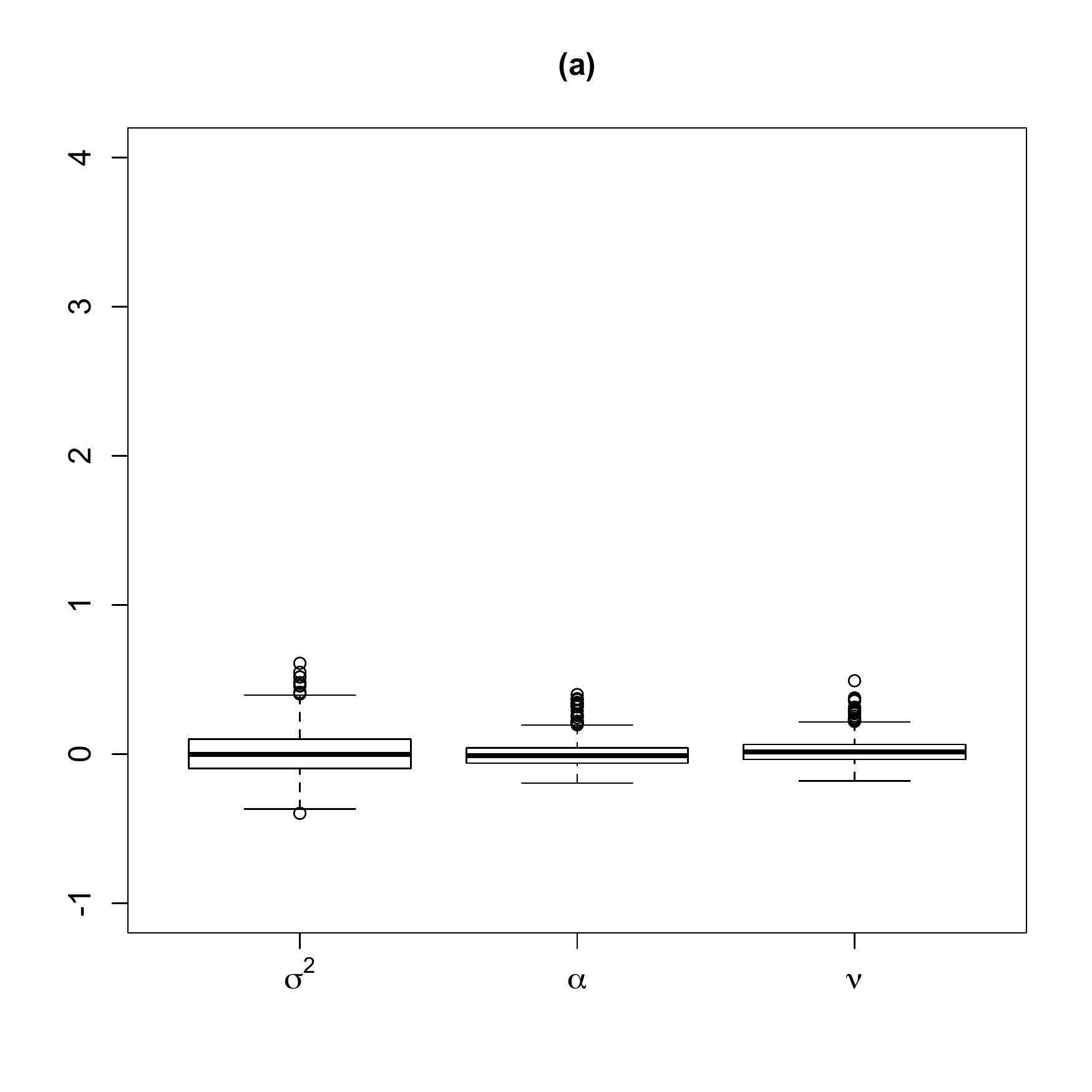} \includegraphics[scale=0.35]{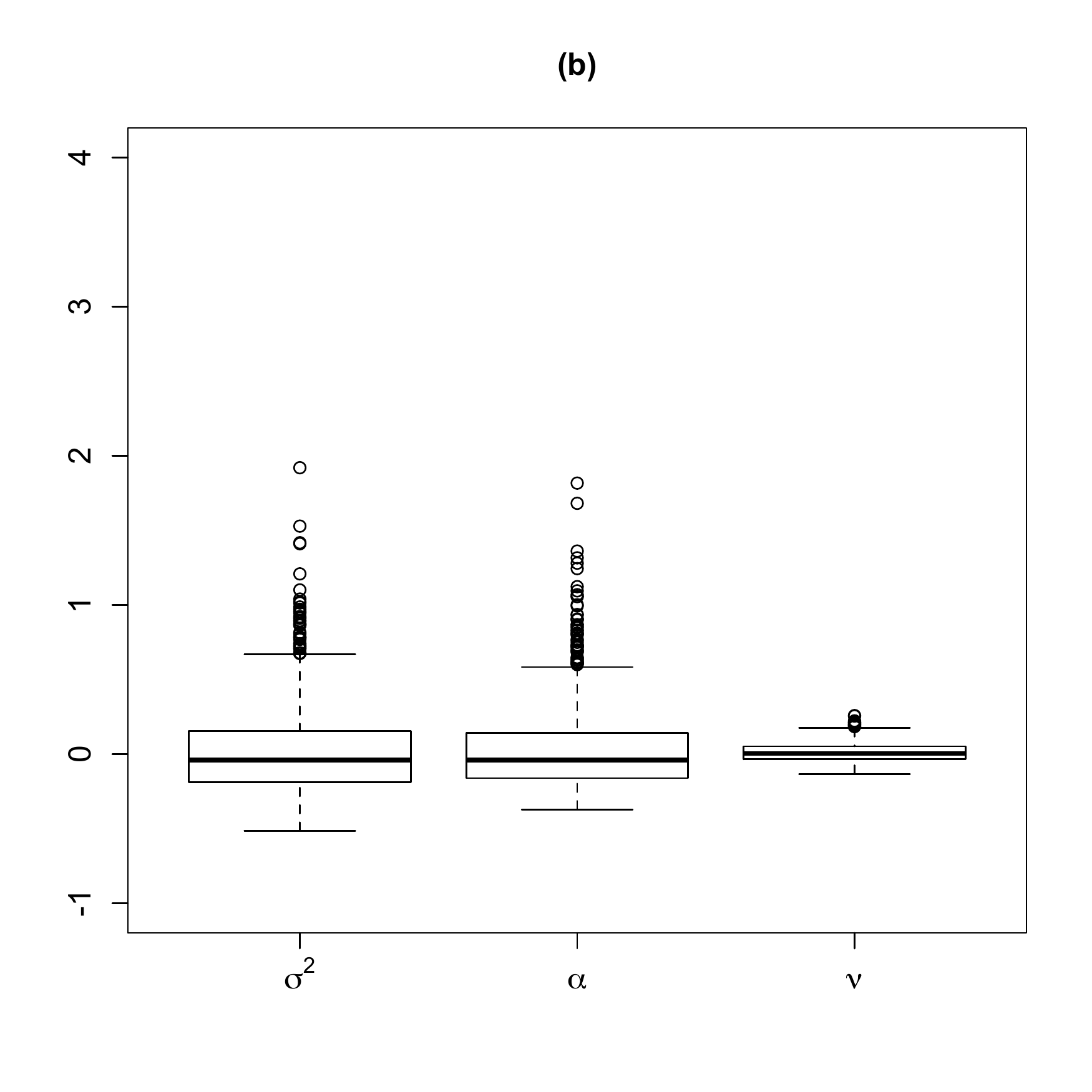}
\includegraphics[scale=0.35]{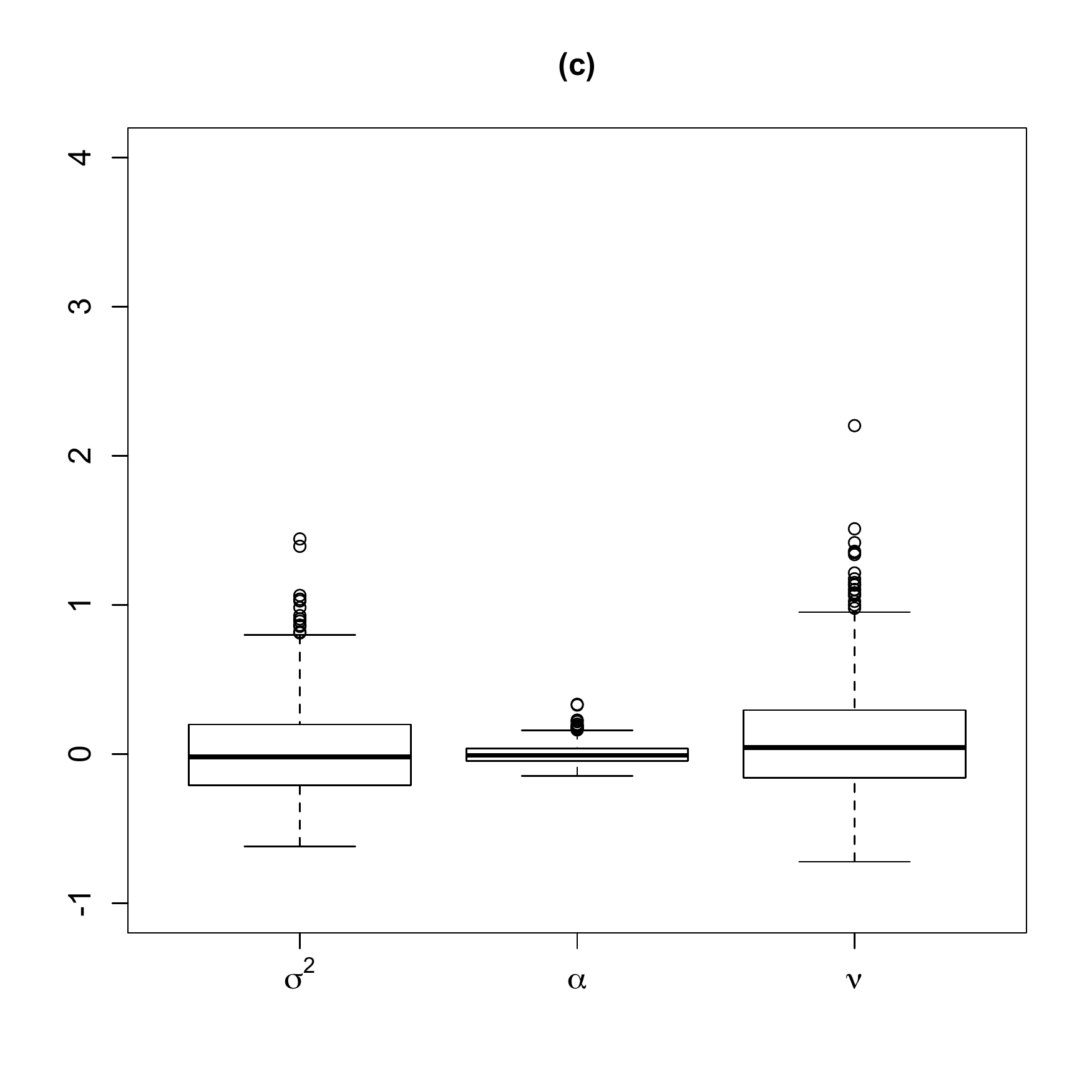}\includegraphics[scale=0.35]{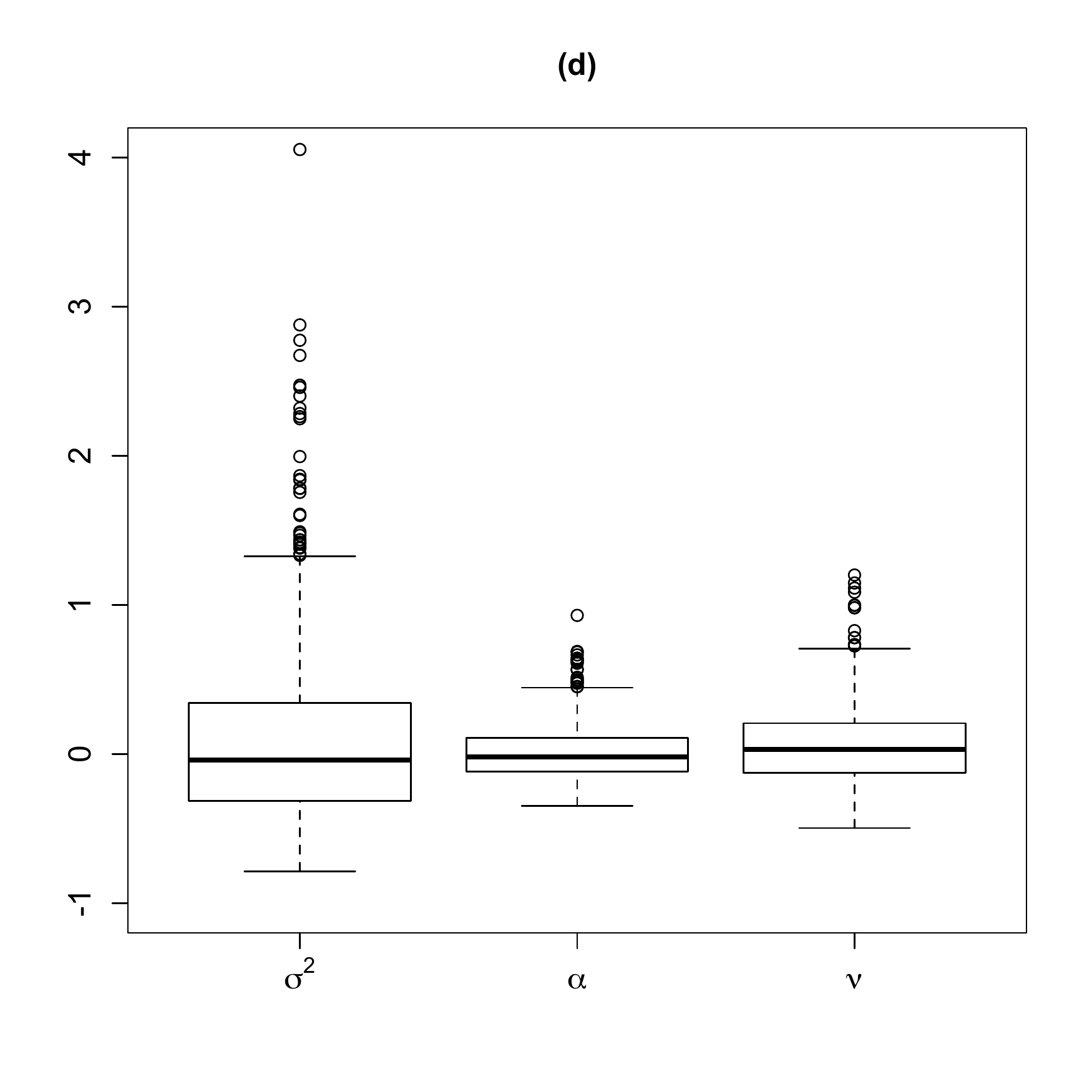}
\caption{Centered boxplots of the ML estimates for the $\mathcal{F}$-Family under Scenarios (a)-(d), based on 1000 independent replications.}
\label{boxplots_ml}
\end{figure}

Figure \ref{boxplots_ml} reports the centered boxplots of the ML estimates under Scenarios (a)-(d), based on 1000 independent replications. Larger values of $\alpha$ and $\nu$ correspond to higher variabilities, but there is no evidence of significant bias.

\subsection{Microergodic parameter}

\cite{zhang} has shown that, for the Mat\'ern class of covariance functions as in Equation (\ref{matern}), not all parameters can be estimated consistently under infill asymptotics.   However, using
the parameterisation analogous to ours, in which $\sigma^2$ is the
variance and $\nu$ determines the degree of mean square differentablity of the random field,
the ML estimator of the
{\it microergodic parameter} $\varpi=\sigma^2 / \alpha^{2\nu}$  is consistent.

\begin{figure}
\centering
\includegraphics[scale=0.35]{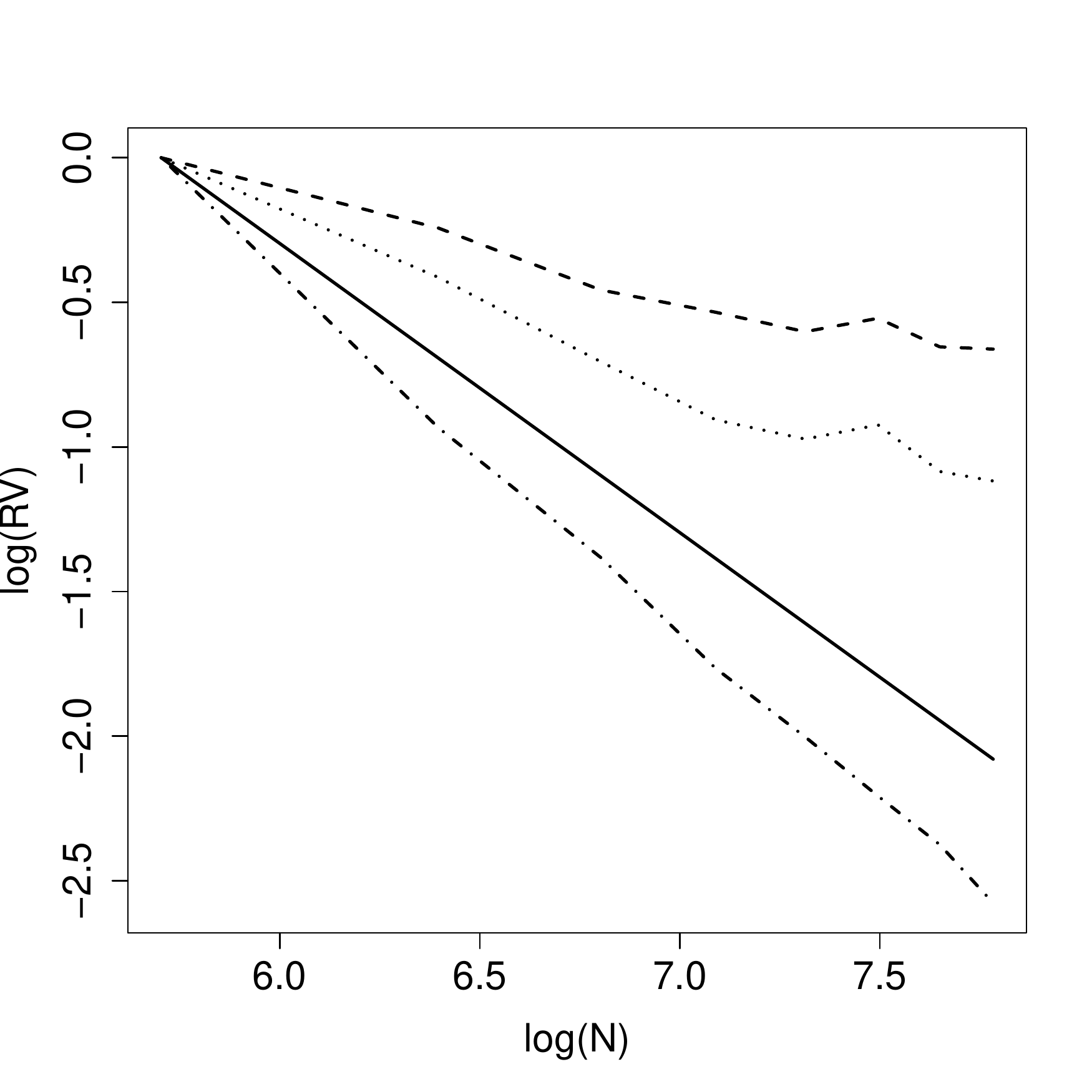}
\caption{Relationship between log-transformed relative variance and sample size from our simulation experiment. Dashed, dotted and dash-dotted lines show the $\log(RV)$ for $\sigma$,$\alpha$ and $\varpi$, from the ${\cal F}$ family. The full line, added as a reference, is a line with slope -1.} \label{fig:new}
\end{figure}

To mimic an infill asymptotic scheme, for each scenario
we now
generate 1000 realisations of $Z$ 
at $N = 300, 600, 900, 1200, 1500, 1800, 2100$ and $2400$ locations 
 uniformly distributed
on the unit sphere, with parameter values
$\sigma^2=1$, $\alpha=0.2$ and $\nu=1/2$.  Table \ref{consistencia} 
summarises the properties of the ML estimates by their emprical
bias and {\it relative variance}, {\em i.e.}, the ratio between their sample variance at each value of $N$ and their sample variance when $N=300$.  The biases are again negligible. 
The standard asymptotic result for parameter estimation is that
 the variance of an ML estimator is 
proportional to $N^{-1}$,
hence the logarithm of the relative variance is linear in $\log(N)$ with
slope $-1$.
Figure \ref{fig:new} shows the empirical relationship between log-transformed relative variance and sample size from our simulation
experiment. For the microergodic parameter $\varpi$, the relationship is close to linear, with estimated slope $-1.048$, whereas for $\sigma^2$ and 
$\alpha$, the estimated slopes are $-0.872$ and $-0.913$, respectively. Also, as $N$ approaches 2400, there is at least a hint that the linearity is breaking down.
 
In summary, the experiment suggests that ML estimates for the parameters of the $\mathcal{F}$-family behave similarly to those of the
planar Mat{\'e}rn model under infill asymptotics. 

\begin{table}
\caption{Bias and Relative Variance (RV) for the maximum likelihood estimates of $\sigma^2$, $\alpha$ and
$\varpi =\sigma^2 / \alpha^{2\nu}$ versus sample size.}
\label{consistencia}
\centering
\begin{tabular}{ccccccccccc}\hline \hline
              & &  \multicolumn{2}{c}{$\sigma^2$}  &&  \multicolumn{2}{c}{$\alpha$} &&  \multicolumn{2}{c}{$\varpi$}\\  \cline{3-4} 
              \cline{6-7} \cline{9-10}
Sample Size       &  &    Bias & RV  &&    Bias & RV &&    Bias & Rel. Var.\\  \hline
300                      & & $-0.00511$ &    1                 &&$-0.00041$        &   1         &&$\,\,\,\,0.04908$&   1         \\
600                      & & $-0.00044$ &   0.784           &&$\,\,\,\,0.00145$&   0.661   &&$-0.00914$&   0.393      \\
900                      & & $-0.00446$ &   0.633           &&$-0.00005$       &  0.494     &&$-0.00183$&   0.251      \\
1200                    & & $-0.00147$ &   0.586           &&$\,\,\,\,0.00042$ &  0.404    &&$-0.00516$&   0.171     \\
1500                    & & $\,\,\,\,0.00029$  &   0.548   &&$\,\,\,\,0.00052$&   0.378    &&$-0.00054$& 0.135    \\
1800                    & & $\,\,\,\,0.00047$ &   0.574   &&$\,\,\,\,0.00073$  &     0.397  &&$-0.00537$& 0.110     \\
2100                    & & $\,\,\,\,0.00174$  &   0.520   &&$\,\,\,\,0.00097$ &    0.338    &&$-0.00740$&  0.093     \\
2400                    & & $-0.00142$ &   0.516            &&$\,\,\,\,0.00009$ &     0.327  &&$-0.00228$&  0.076    \\   \hline \hline
\end{tabular}
\end{table}

\section{Data illustration} \label{sec5}

We illustrate the predictive performance of the ${\cal F}$ class of covariance functions on a dataset of Precipitable Water Content (PWC) in ${\rm kg}/{\rm m}^2$, donwloadable at $www.esrl.noaa.gov$. 
This data-product is considered to be representative of the state of the Earth system \citep{kalnay1996ncep} and has been used in regional studies  of seasonal stream flow and  water scarcity \citep{muller2014analytical, muller2016comparing}. Here, we analyse the $2017$ average of PWC on a grid with spacing $2.5^\circ$ degrees of longitude and latitude.  

\begin{figure}
\begin{subfigure}{.5\linewidth}
\centering
\includegraphics[scale=0.22]{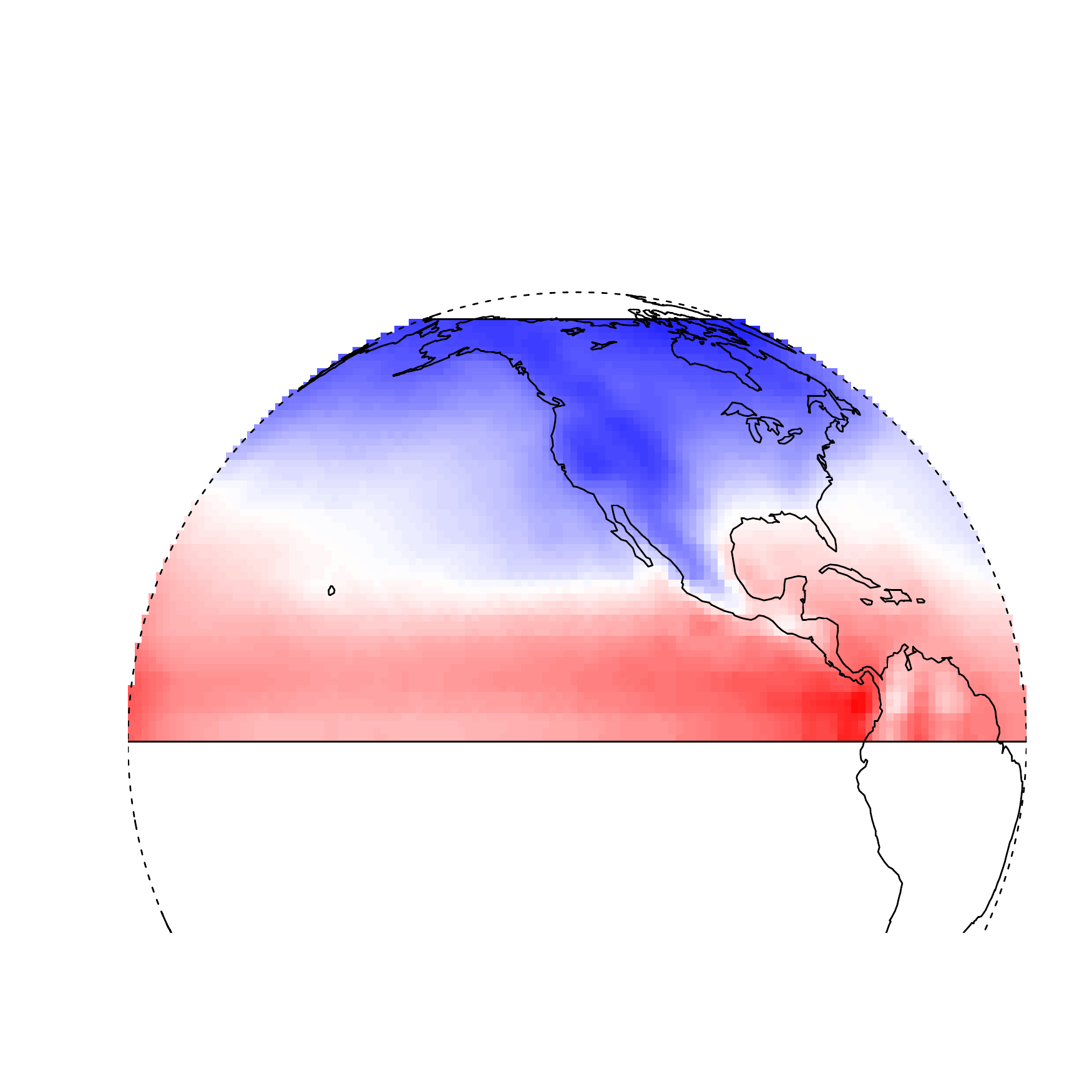}
\caption{Longitude $-120^{\circ}$.}
\label{fig:sub_dataset1}
\end{subfigure}%
\begin{subfigure}{.5\linewidth}
\centering
\includegraphics[scale=0.22]{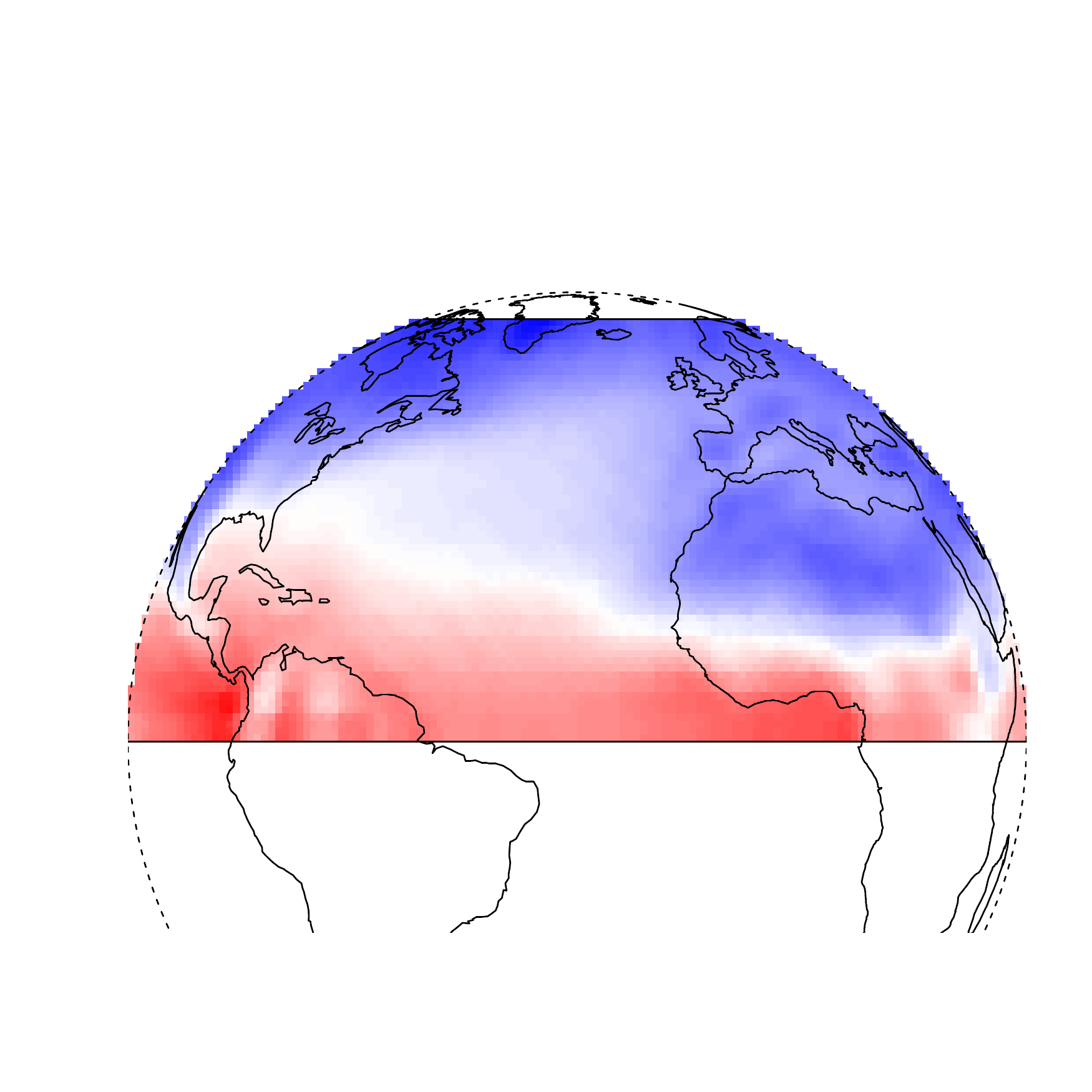}
\caption{Longitude $-30^{\circ}$.}
\label{fig:sub_dataset2}
\end{subfigure}\\[1ex]
\begin{subfigure}{.5\linewidth}
\centering
\includegraphics[scale=0.22]{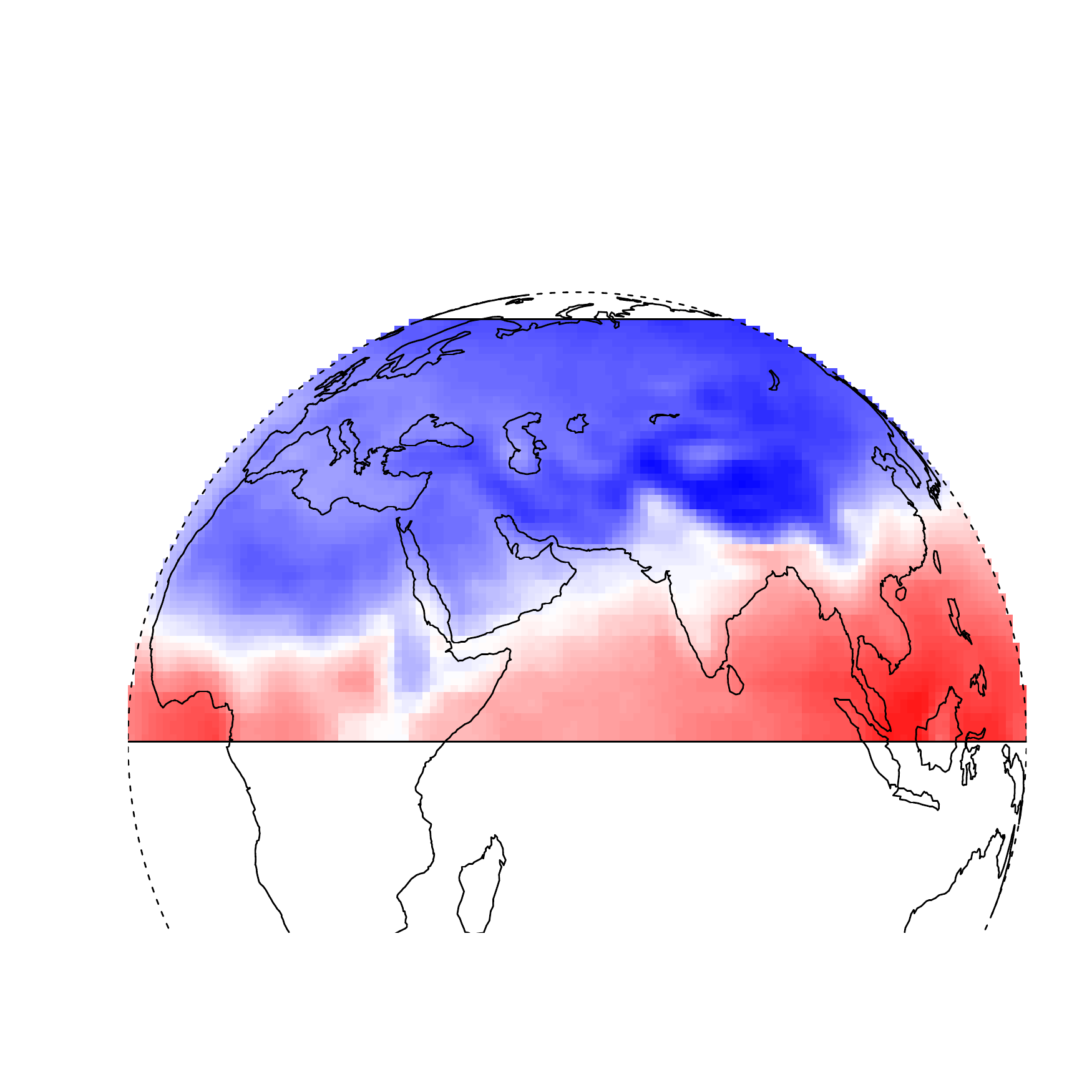}
\caption{Longitude $60^{\circ}$.}
\label{fig:sub_dataset3}
\end{subfigure}%
\begin{subfigure}{.5\linewidth}
\centering
\includegraphics[scale=0.22]{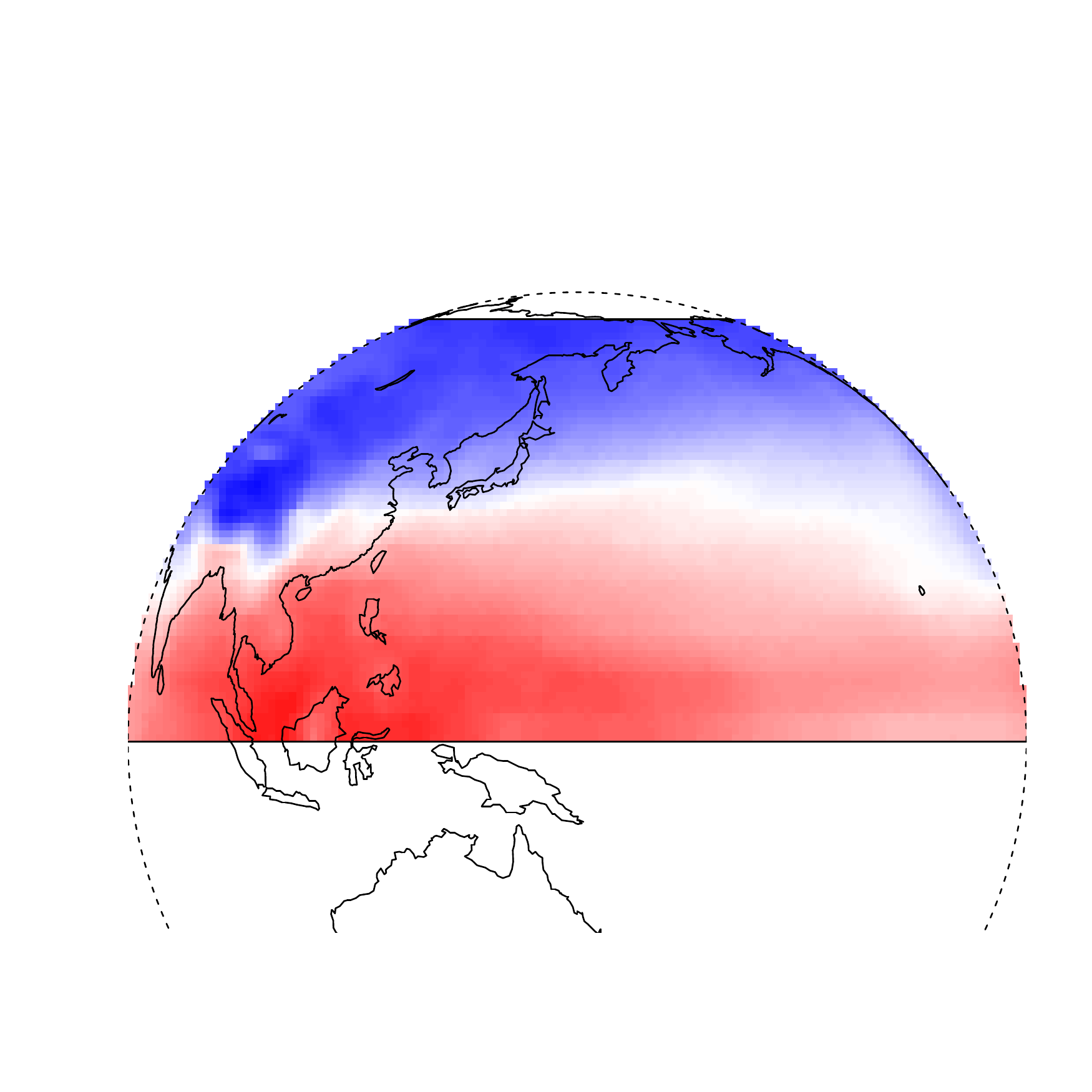}
\caption{Longitude $150^{\circ}$.}
\label{fig:sub_dataset4}
\end{subfigure}\\[1ex]
\begin{subfigure}{\linewidth}
\centering
\includegraphics[scale=0.85]{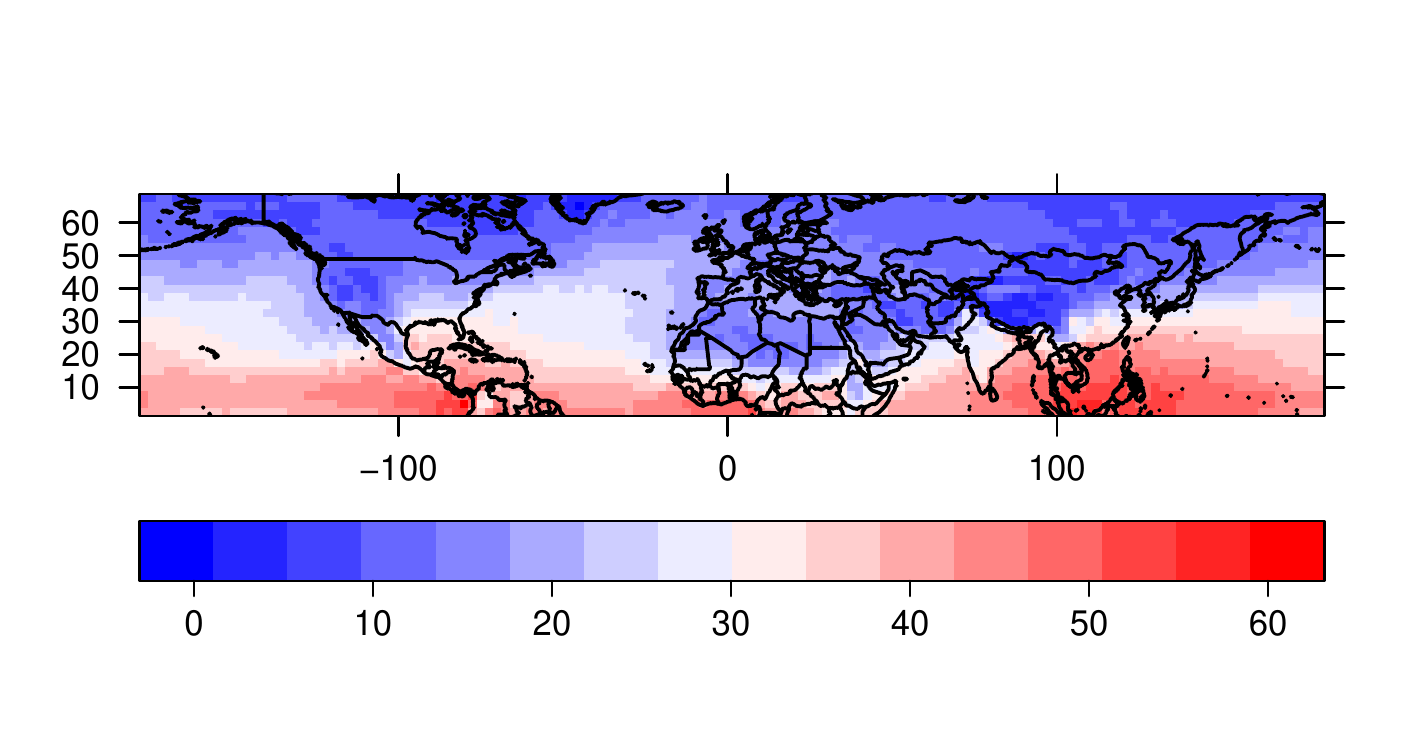}
\caption{Map projection of the data.}
\label{fig:sub_dataset_flat}
\end{subfigure}
\caption{Plot of the data over planet Earth seen from different longitudes \eqref{fig:sub_dataset1}--\eqref{fig:sub_dataset4}.  The whole data is depicted in \eqref{fig:sub_dataset_flat} through a map projection on the plane. The same scale was used for all panels. An animated version is available at  \texttt{https://github.com/FcoCuevas87/F${}\_{}$Family${}\_{}$cova}. }
\label{fig:data_set}
\end{figure}

\begin{figure}
\begin{subfigure}{.5\linewidth}
\centering
\includegraphics[scale=0.35]{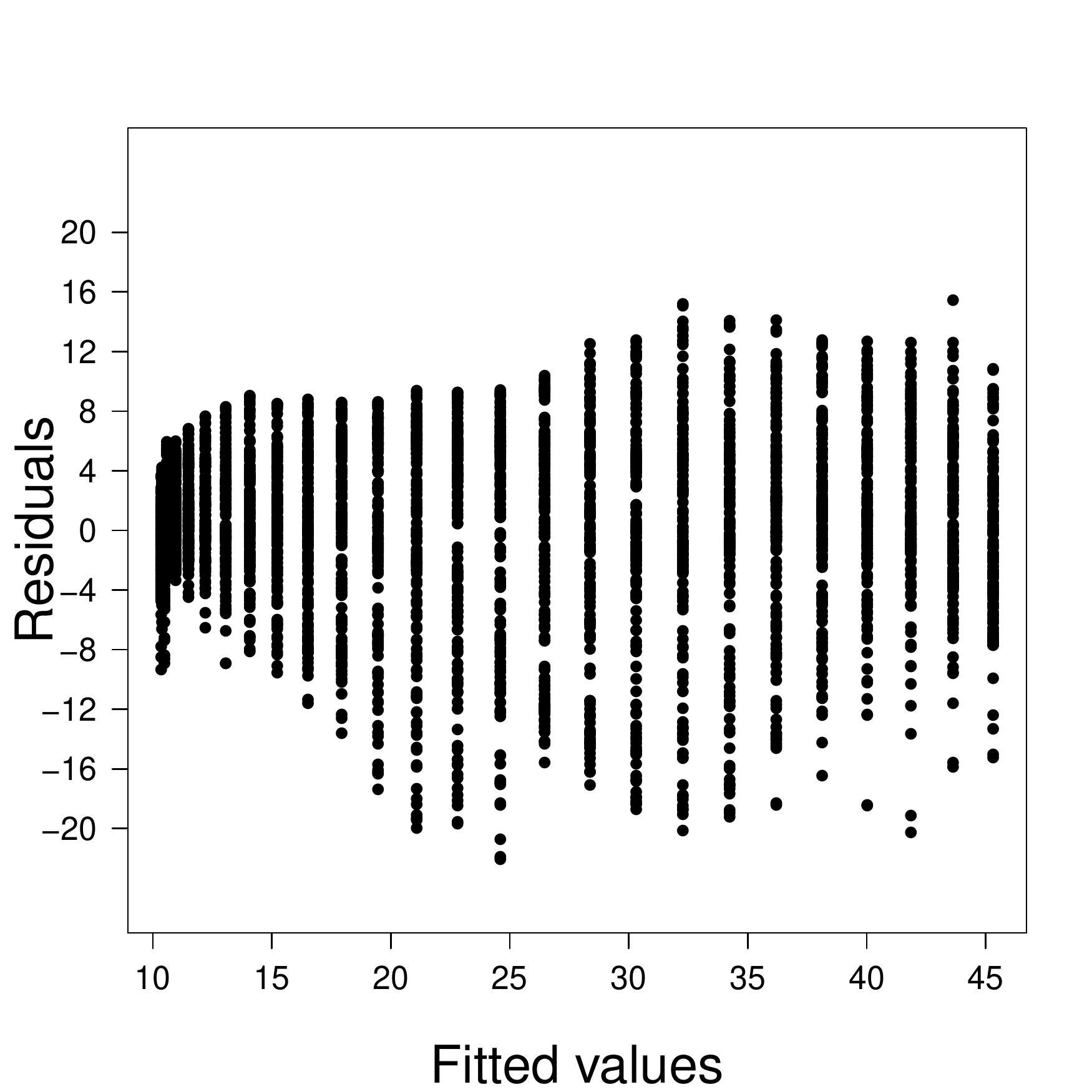}
\caption{Constant variance.}\label{fig:sub_residuals1}
\end{subfigure}%
 \begin{subfigure}{.5\linewidth}
\centering
\includegraphics[scale=0.35]{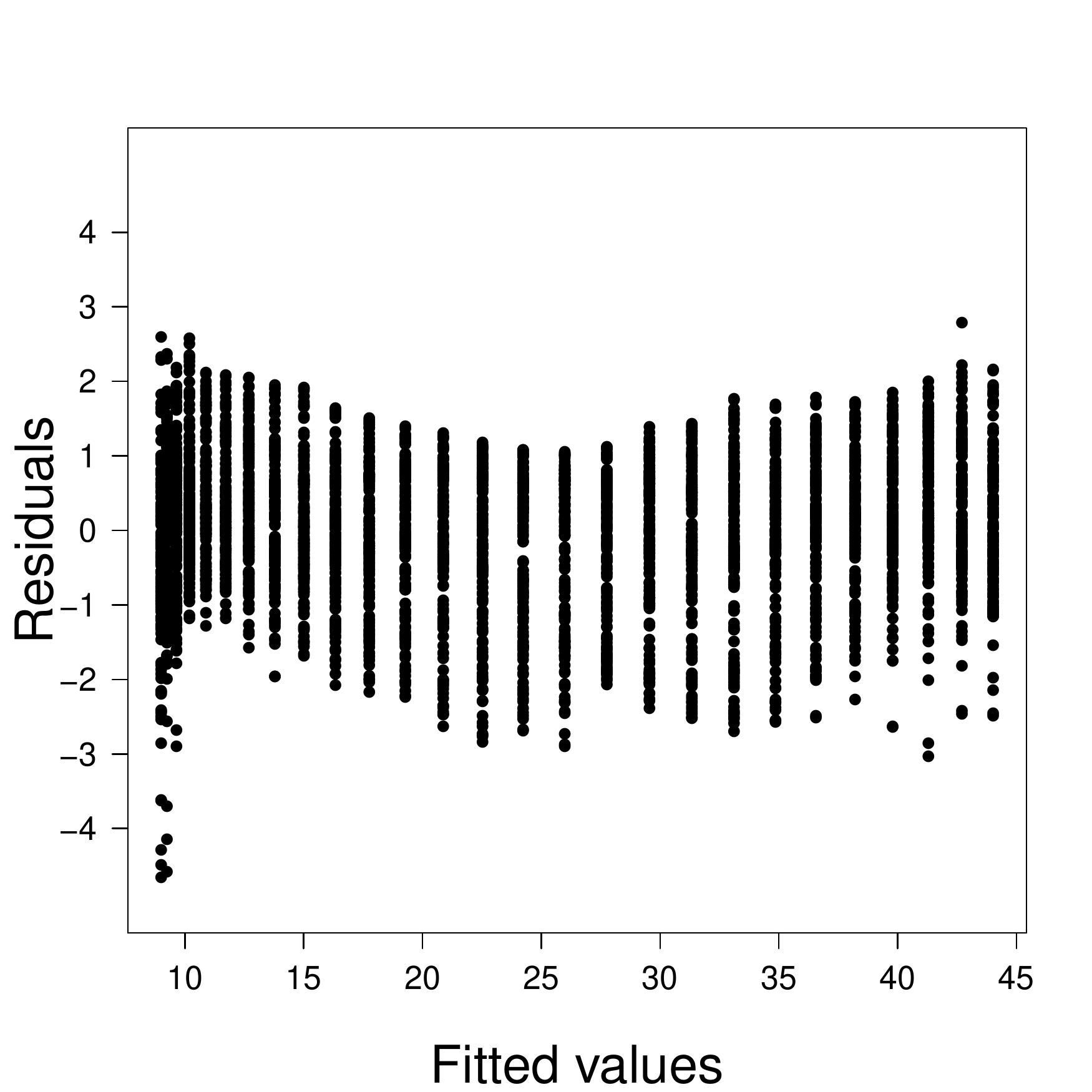}
\caption{Non-constant variance}\label{fig:sub_residuals2}
\end{subfigure}
\caption{Residuals obtained from the regression with constant \eqref{fig:sub_residuals1} and non-constant \eqref{fig:sub_residuals2} variance.}\label{fig:residuals}
\end{figure}

\begin{figure}
\begin{subfigure}{.5\linewidth}
\centering
\includegraphics[scale=0.22]{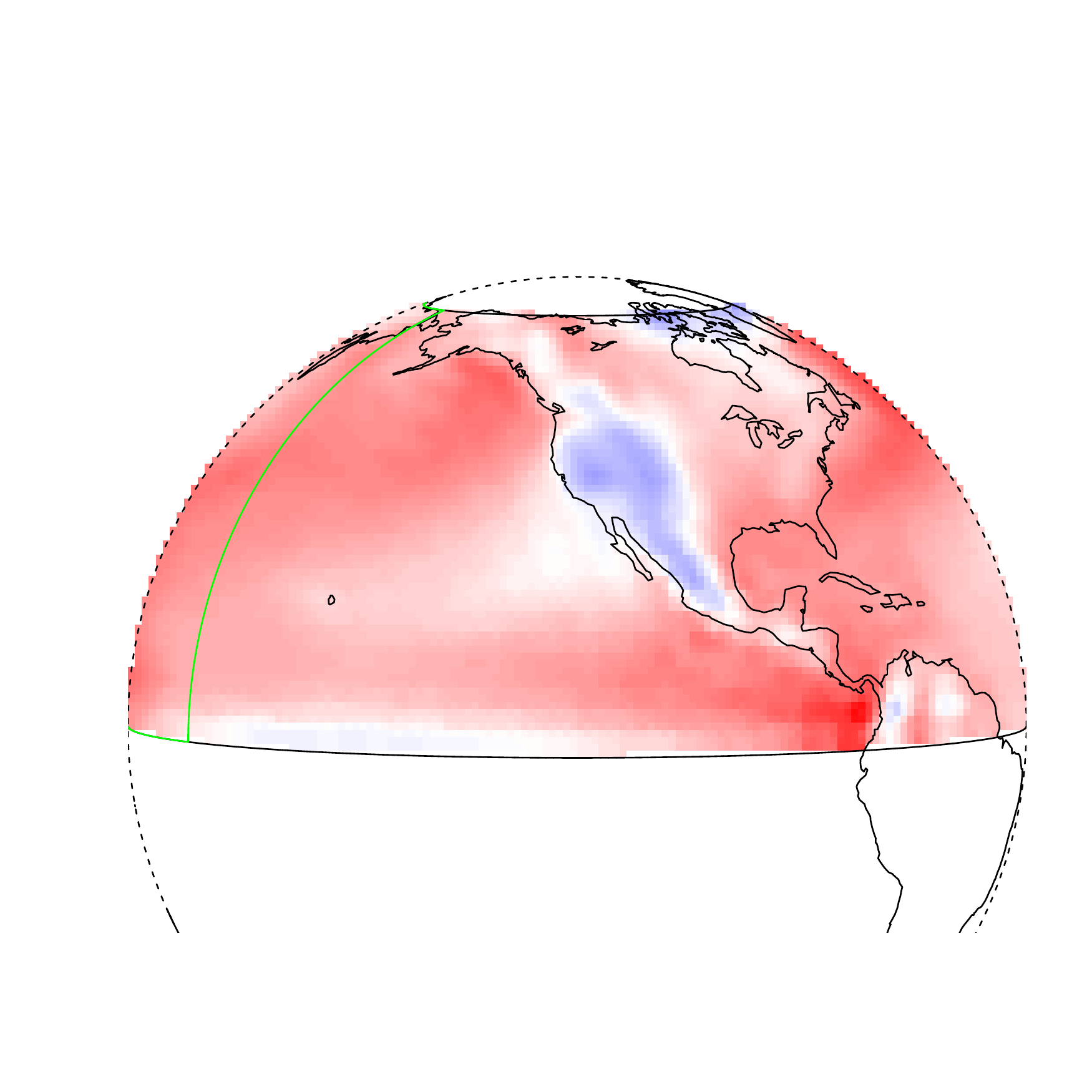}
\caption{Longitude $-120^{\circ}$.}
\label{fig:sub_filter1}
\end{subfigure}%
\begin{subfigure}{.5\linewidth}
\centering
\includegraphics[scale=0.22]{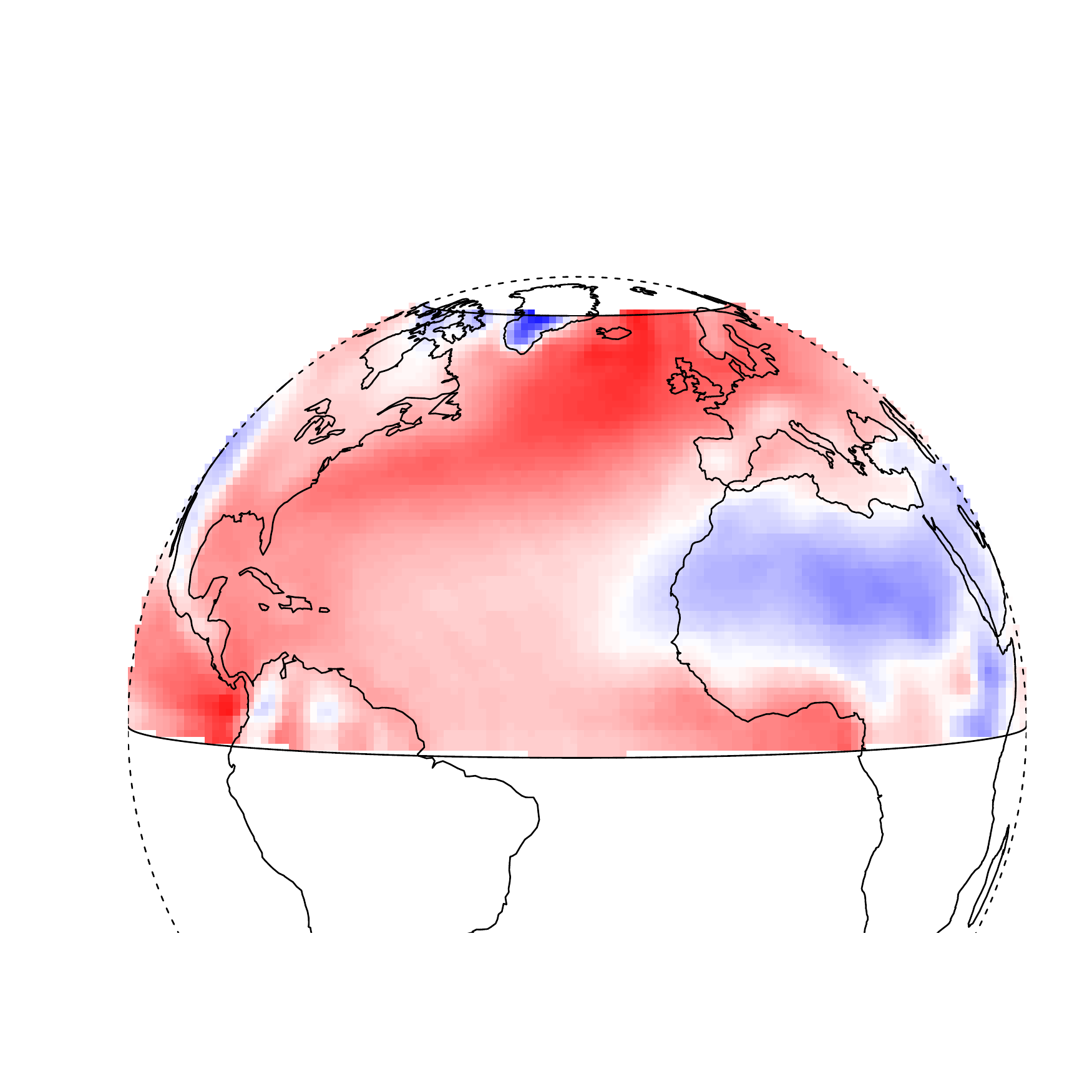}
\caption{Longitude $-30^{\circ}$.}
\label{fig:sub_filter2}
\end{subfigure}\\[1ex]
\begin{subfigure}{.5\linewidth}
\centering
\includegraphics[scale=0.22]{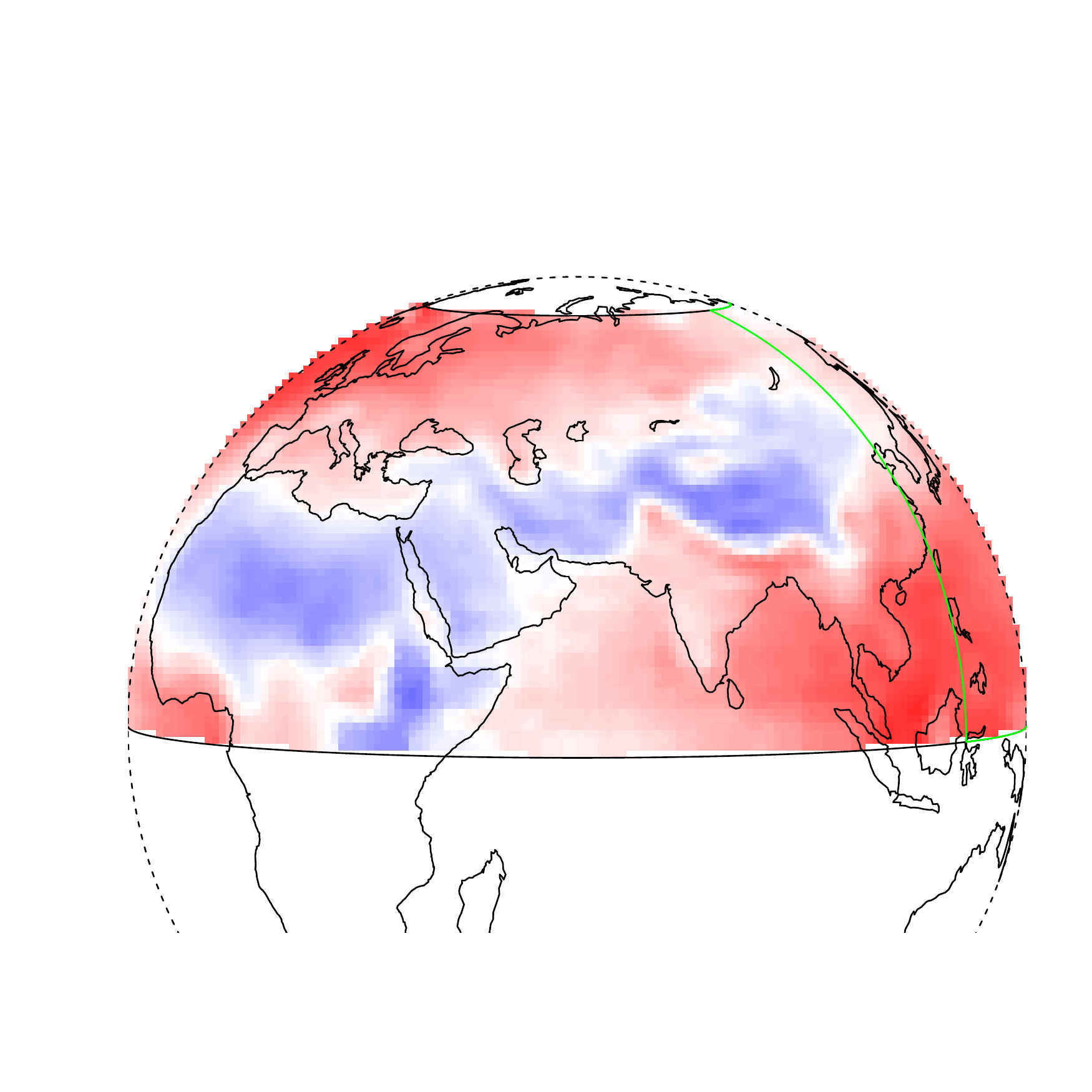}
\caption{Longitude $60^{\circ}$.}
\label{fig:sub_filter3}
\end{subfigure}%
\begin{subfigure}{.5\linewidth}
\centering
\includegraphics[scale=0.22]{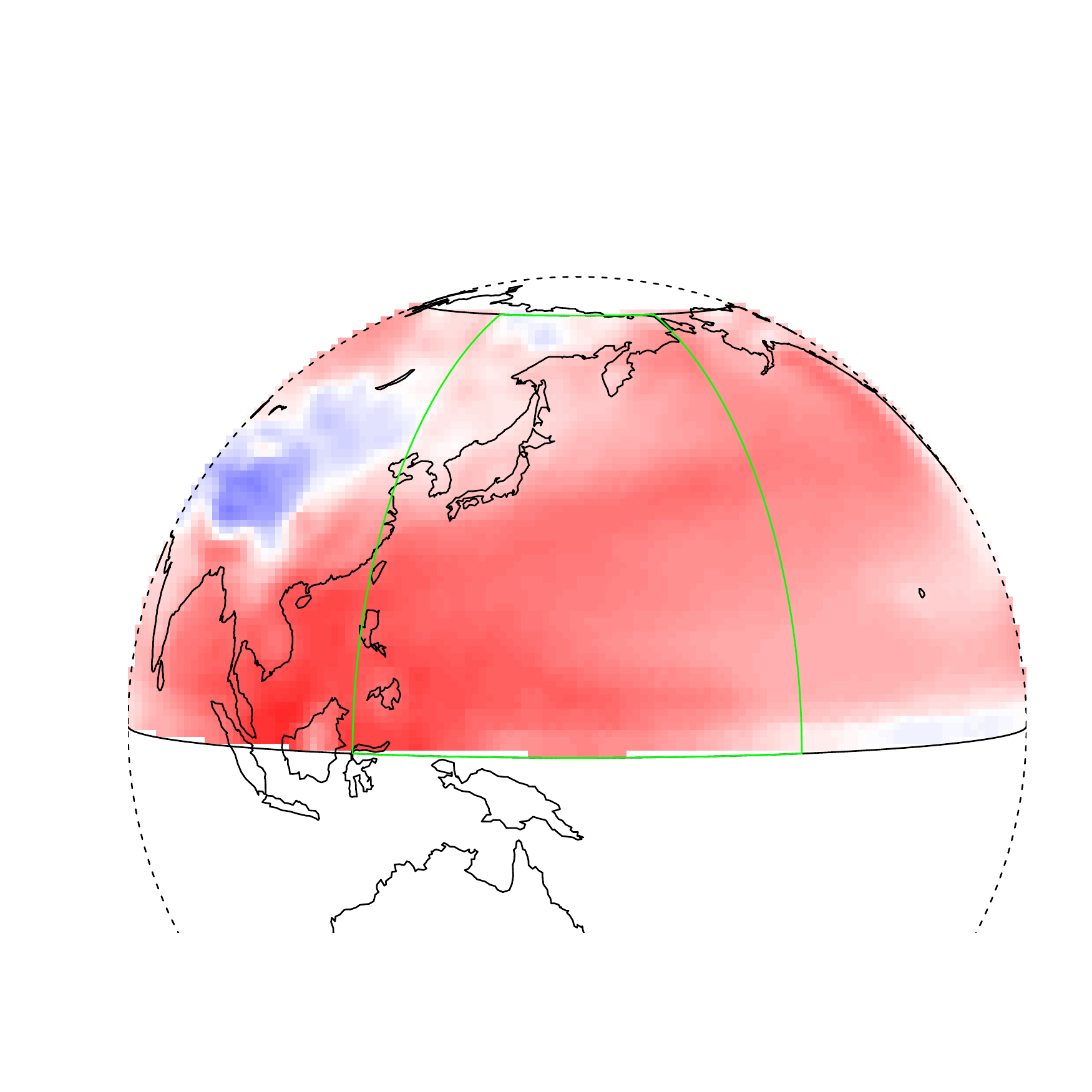}
\caption{Longitude $150^{\circ}$.}
\label{fig:sub_filter4}
\end{subfigure}\\[1ex]
\begin{subfigure}{\linewidth}
\centering
\includegraphics[scale=0.85]{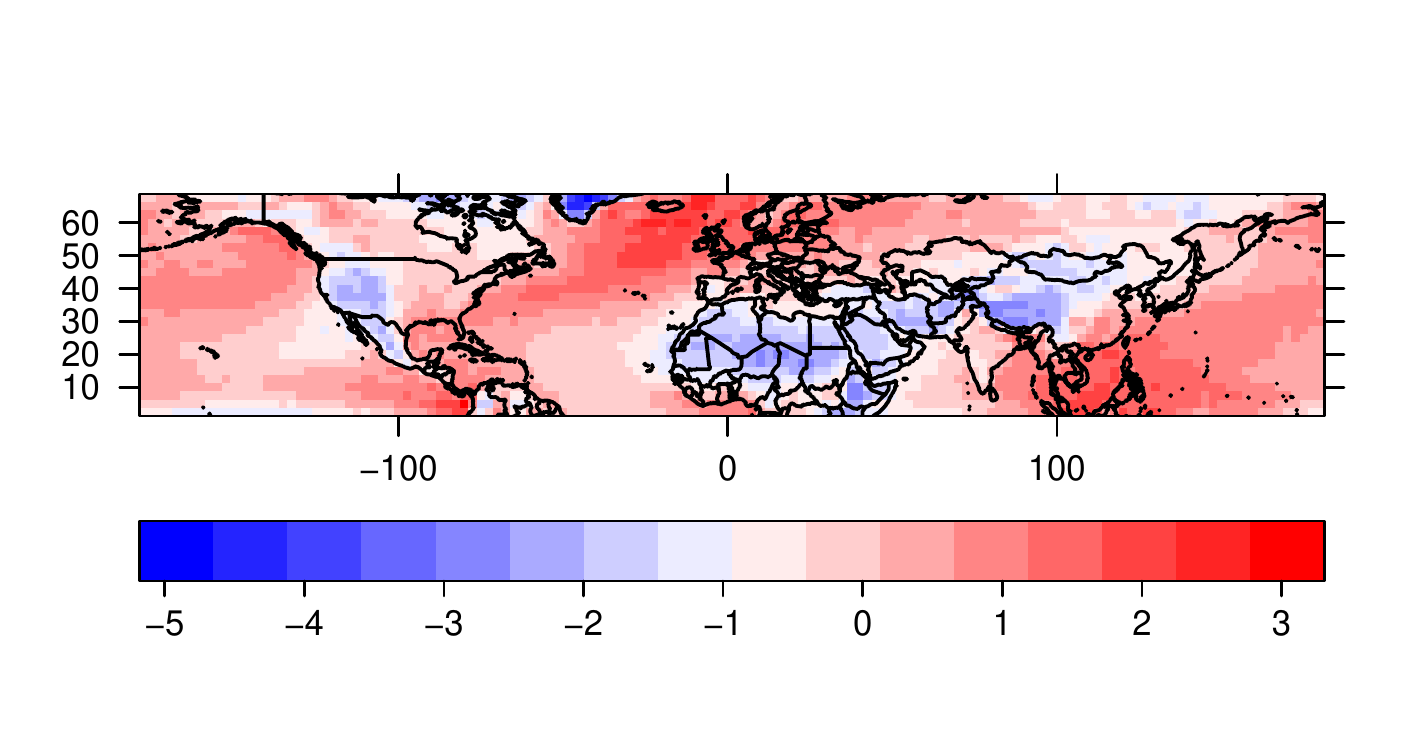}
\caption{Map projection of the filtered data.}
\label{fig:sub_filter_flat}
\end{subfigure}
\caption{Plot of the filtered data on planet Earth seen from different longitudes \eqref{fig:sub_filter1}--\eqref{fig:sub_filter4}.  The whole set of residuals is then depicted in \eqref{fig:sub_filter_flat} through a map projection on the plane. The same scale was used for all panels. The green region of panel \eqref{fig:sub_filter4} is the window used as validation set. An animated version is available at \texttt{https://github.com/FcoCuevas87/F${}\_{}$Family${}\_{}$cova}. }
\label{fig:test}
\end{figure}

Our study focuses on the region within latitudes $0^{\circ}$ and $70^{\circ}$ North, in order to mitigate the effect of non-stationarities over southern latitudes \citep{ste07} and to avoid numerical instabilities around the North Pole \citep{castruccio1}. The data is shown in Figure \ref{fig:data_set}, where we can observe a trend that depends on latitude. An animation is also available (see  the caption of Figure \ref{fig:data_set}). We remove the spatial trend through a simple harmonic regression model,
\begin{equation}\label{mean_function}
 \mu(\bm{x}) = {\rm E} \left (Z(\bm{x})  \right) = \eta_0  + \eta_1 \cos\left( \frac{\pi L}{90^\circ}  \right) + \eta_2 \sin\left( \frac{\pi L}{90^\circ} \right),
\end{equation}
where $L$ denotes the latitude of the point $\bm{x}$, in degrees. Estimation was done by least squares, obtaining $\widehat{\boldsymbol{\eta}} = (\widehat{\eta}_{0},\widehat{\eta}_{1},\widehat{\eta}_{2}) = (32.89, 14.02, -17.65)$. Figure \ref{fig:sub_residuals1} shows the scatterplot of fitted values versus residuals, which suggests heteroscedasticity of the data. We then follow \citet{verbyla1993modelling} and consider a log-linear model for the standard deviation:
\begin{equation} \label{sigma_function}
\log(\sigma(L,\ell)) = \gamma_0  + \gamma_1 \cos\left( \frac{\pi L}{90^\circ}  \right) + \gamma_2 \sin\left( \frac{\pi L}{90^\circ} \right) + \gamma_{3}\sin\left( \frac{\pi L}{90^\circ} \right)\sin\left( \frac{\pi \ell}{90^\circ} \right),
\end{equation}
where $L$ and $\ell$ denote longitude and latitude, respectively. Next, we estimate the parameters of $\mu(\cdot)$ and $\sigma(\cdot,\cdot)$ using the procedures detailed in \citet{verbyla1993modelling} obtaning $\widehat{\boldsymbol{\eta}} = (29.43,15.80,-13.09)$ and $(\widehat{\gamma}_{0},\widehat{\gamma}_{1},\widehat{\gamma}_{2},\widehat{\gamma}_{3}) = (0.64,1.00,0.97,0.23)$. Figure \ref{fig:sub_residuals2} shows the scatterplot of fitted values versus residuals based on \eqref{mean_function} and \eqref{sigma_function}, whilst Figure \ref{fig:test} shows the residuals on planet Earth, indicating a good fit of the mean and variance functions.

To model the correlation of the  residuals, we propose the following models:
\begin{enumerate}
\item the $\mathcal{F}$ covariance function, defined  according to Equation \eqref{sergio-ramos}.
\item the circular-Mat\'ern covariance function \citep{guinness} given by \eqref{guinnes}. As explained in Section \ref{sec2}, in practice a truncation of the series expansion is needed. Therefore, we truncate the sum after $1000$ terms. \cite{lang-schwab} adopt the same strategy and give bounds for the approximation in the mean square sense.
\item A Mat\'ern covariance function $\mathcal{M}_{\nu,\alpha}(d_\text{CH})$, where $d_\text{CH}$ denotes the chordal distance.
\end{enumerate} 
For model selection purposes, we compare the performance of the proposed covariance function models with respect to ML estimation and kriging predictions. Specifically, we repeat the following two-step procedure $500$ times:

\begin{itemize}
\item[Step 1] We first sample $200$ data-locations independently at random over the region delimited by latitudes $0^\circ$ to $70^\circ$ and longitudes $-180^\circ$ to $120^\circ$. We use this data as a training set to calculate the ML estimates for each of the three models. 
\item[Step 2] Then we sample, $100$ times, $20$ data-locations in the region delimited by $0^\circ$ to $70^\circ$ in latitude and $120^\circ$ to $180^\circ$ in longitude as a validation set (see Figure \ref{fig:sub_filter4}). 
\end{itemize}

\begin{figure}
\centering
\includegraphics[scale=0.35]{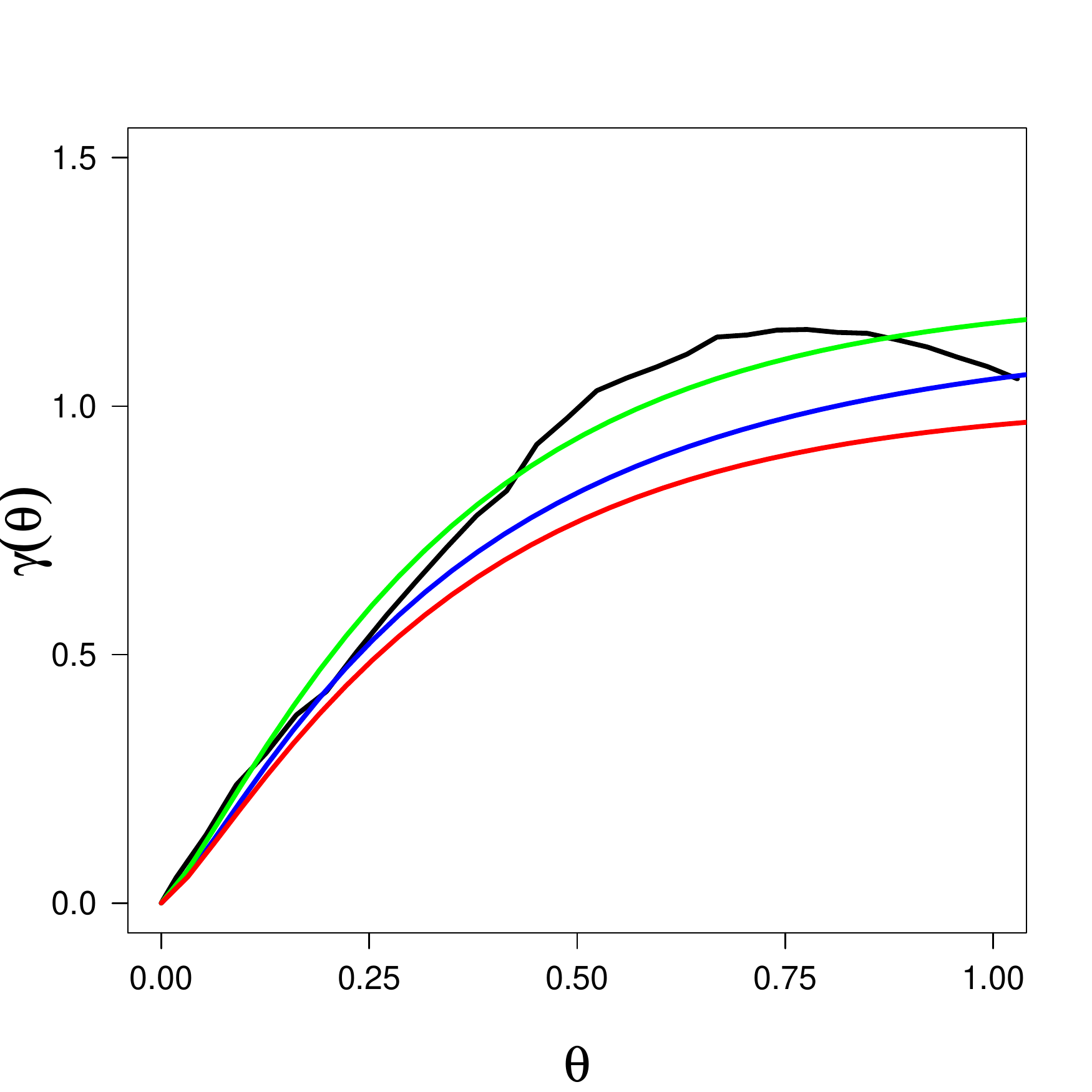}
\caption{Empirical variogram (black) and fitted variograms with $\mathcal{F}$-family (red), chordal Mat{\'e}rn (green) and circular Mat{\'e}rn (blue).}
\label{fig:fitted_variograms}
\end{figure}

A similar experiment has been carried out by \cite{jeong-jun2}, who note that such a scenario may arise in practice when the interest is in predicting over the ocean but most observations are on land. Table \ref{mle} reports the average ML estimates for each model, their empirical standard errors and the maximised log-likelihood. In addition, Figure \ref{fig:fitted_variograms} shows the sample variogram and the three correlation function models using the average ML estimates. In all three cases, the estimate $\hat{\nu}$ corresponds to a continuous but nondifferentiable random field. The values of the maximised log-likelihood are very similar.

\begin{figure}
\begin{subfigure}{\linewidth}
\centering
\includegraphics[scale=0.25]{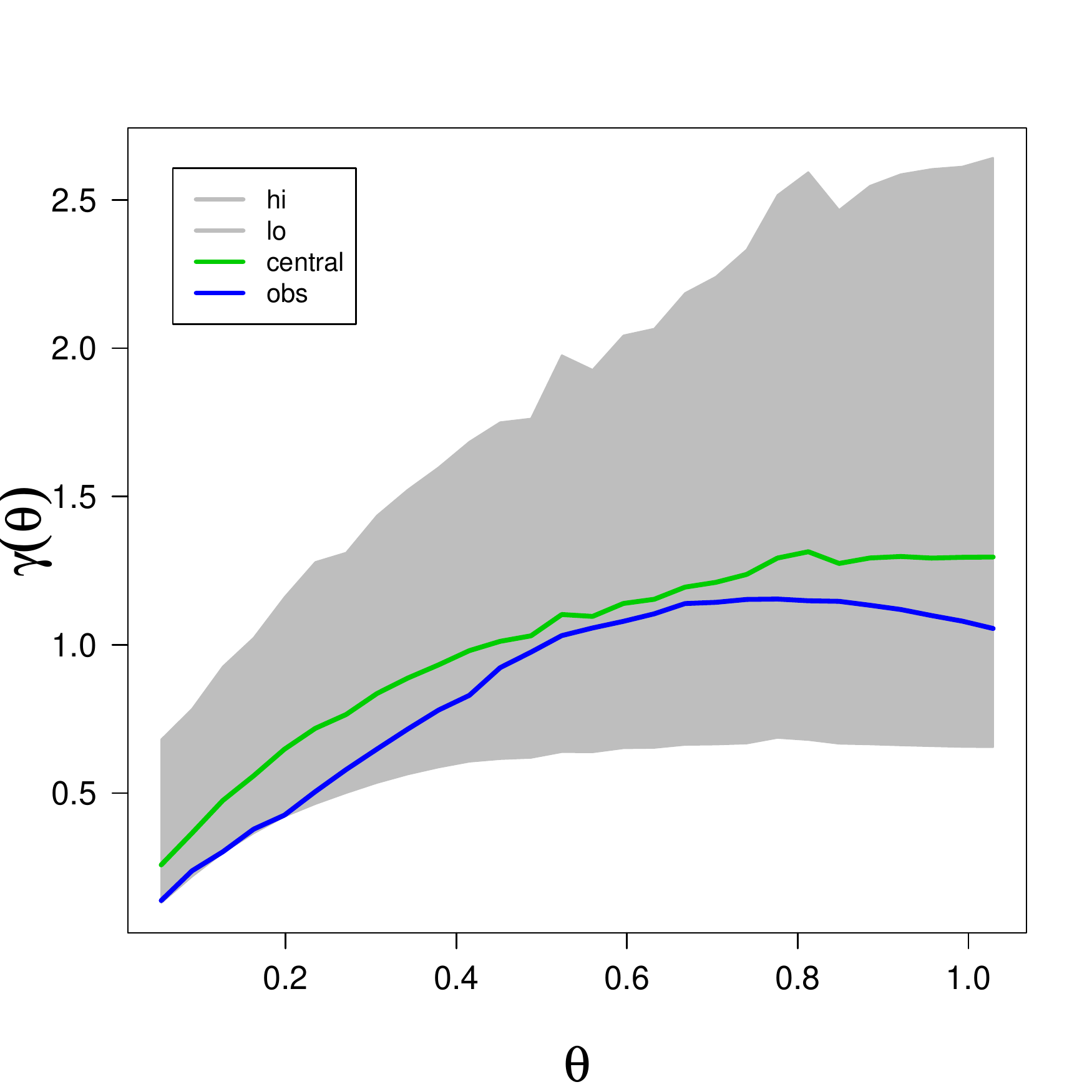}
\caption{Isotropic variogram using all data points. \\ }
\label{fig:sub_emp_variog_iso}
\end{subfigure}
\begin{subfigure}{.45\linewidth}
\centering
\includegraphics[scale=0.25]{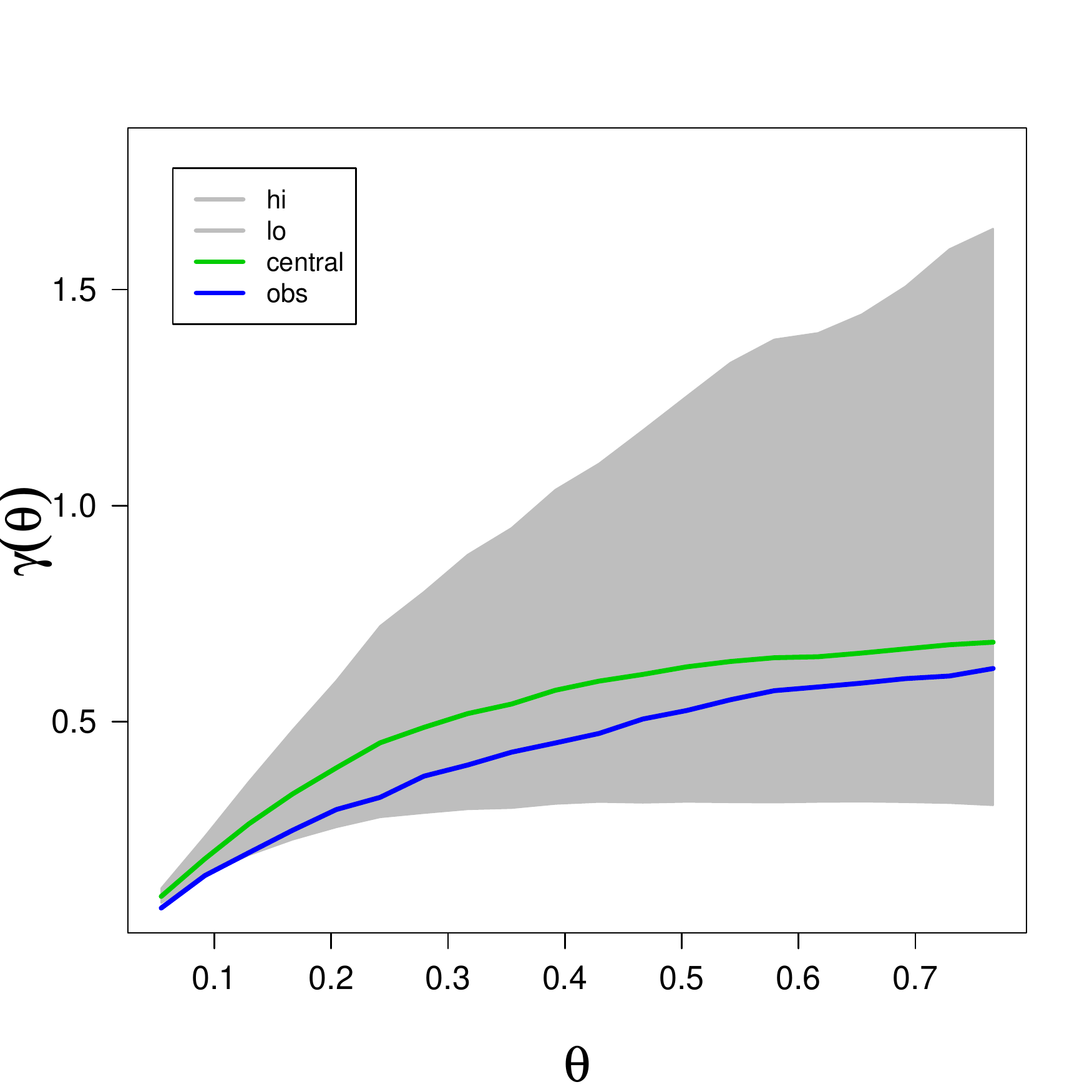}
\caption{Data points at south of latitude $35^{\circ}$ and west of longitude $-1^{\circ}$.}
\label{fig:sub_emp_variog_lat}
\end{subfigure}\hfill
\begin{subfigure}{.45\linewidth}
\centering
\includegraphics[scale=0.25]{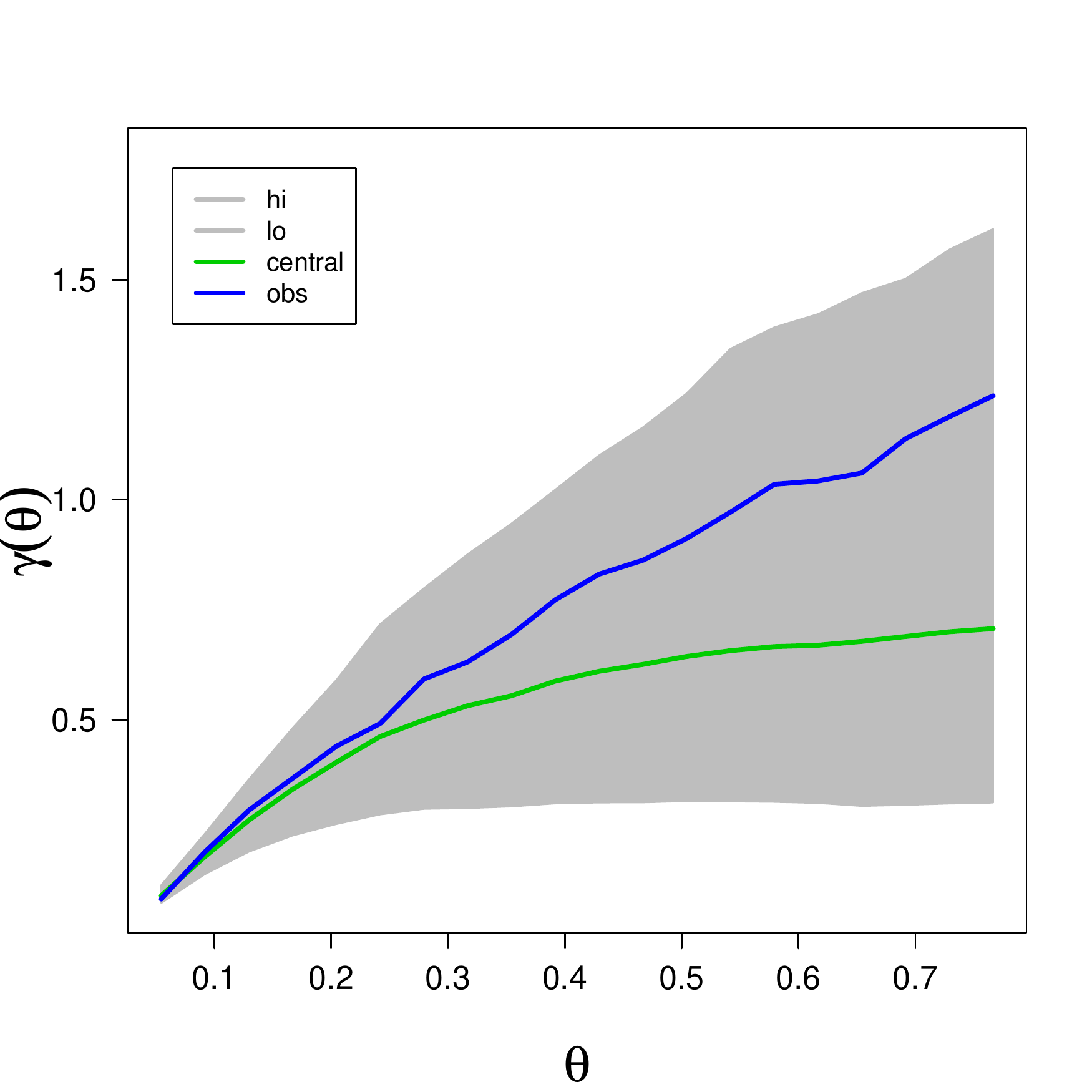}
\caption{Data points at south of latitude $35^{\circ}$ and east of longitude $-1^{\circ}$.}
\label{fig:sub_emp_variog_lon}
\end{subfigure}
\begin{subfigure}{.45\linewidth}
\centering
\includegraphics[scale=0.25]{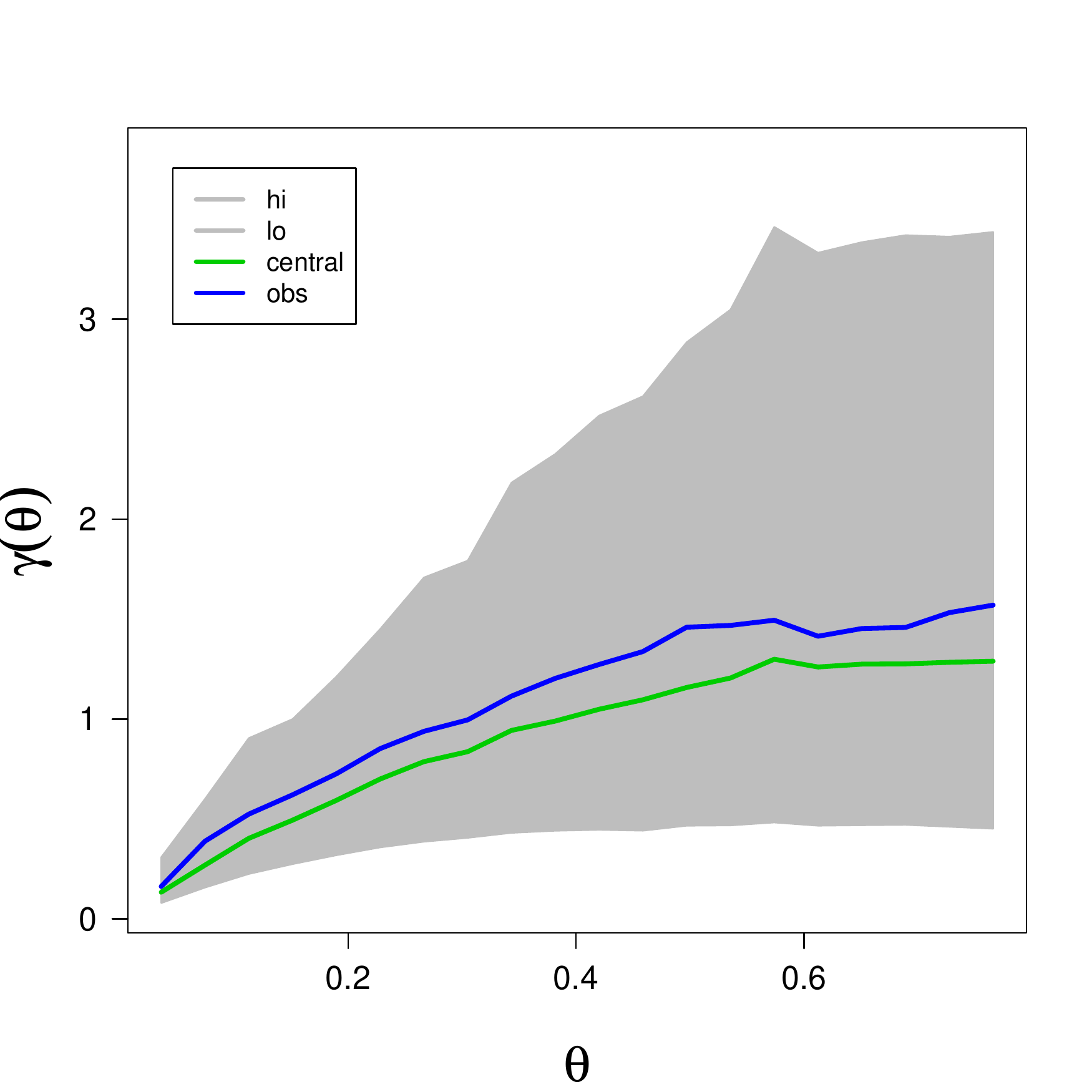}
\caption{Data points at north of latitude $35^{\circ}$ and west of longitude $-1^{\circ}$.}
\label{fig:sub_emp_variog_lon2}
\end{subfigure}\hfill
\begin{subfigure}{.45\linewidth}
\centering
\includegraphics[scale=0.25]{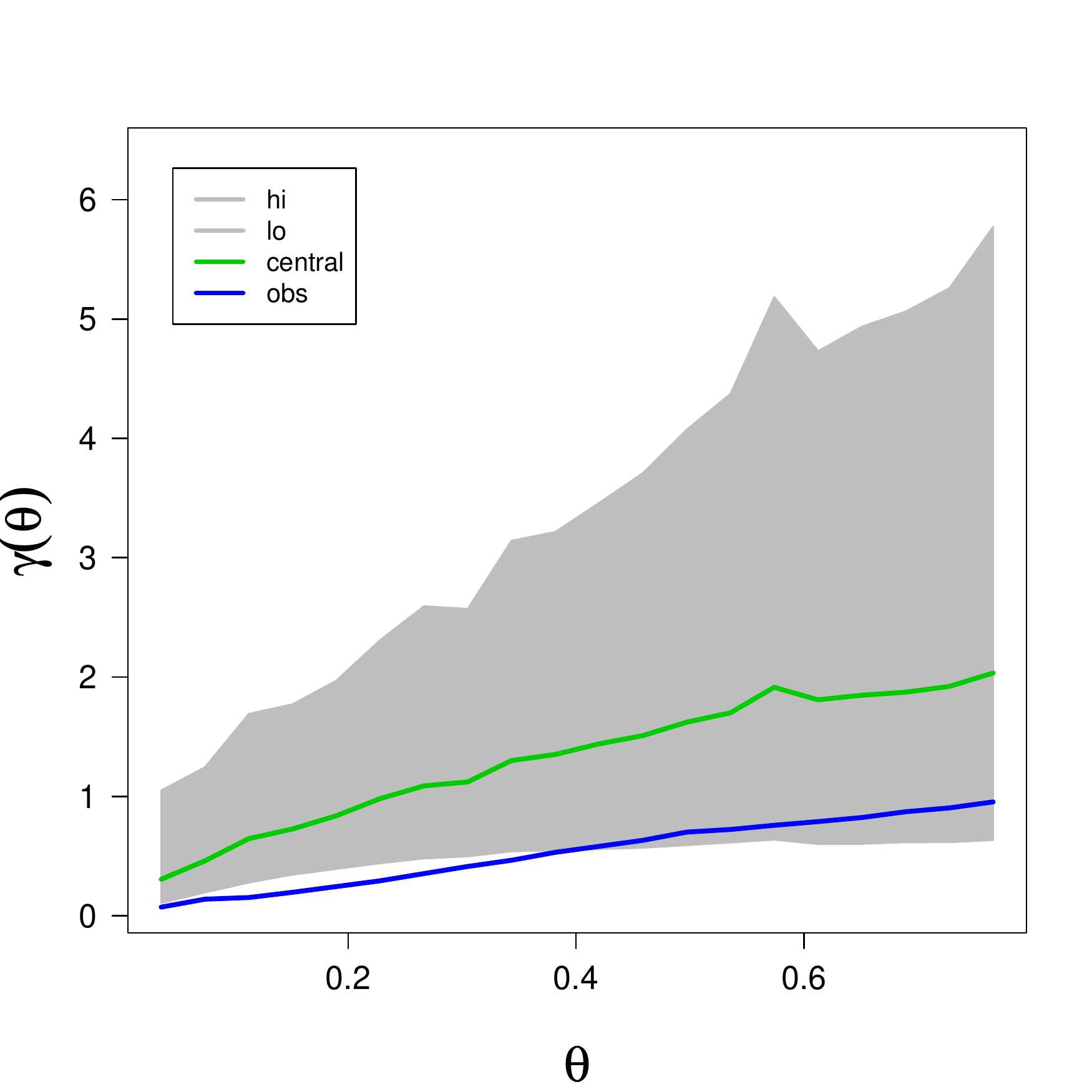}
\caption{Data points at north of latitude $35^{\circ}$ and east of longitude $-1^{\circ}$.}
\label{fig:sub_emp_variog_lon3}
\end{subfigure}
\caption{Empirical semi-variograms and global rank envelopes at 95\% using the $\mathcal{F}$-family model with different subsets of the data (see panels). The different cut points ensure that the number of points is the same on each sub-region. The blue line represents the semi-variogram on each quadrant whilst the green line is the mean curve obtained by simulations.}
\label{fig:emp_variograms}
\end{figure}

We also inspect for anisotropy of the residuals as a model validation procedure. Figure \ref{fig:sub_emp_variog_iso} shows the empirical semi-variogram for the filtered data together with the $2.5\%$, median and $97.5\%$ pointwise quantiles from $2499$  simulations of the fitted ${\cal F}$-family model as described below. In addition, we investigate the isotropy assumption by estimating the semi-variogram on different sub-regions of the part of the planet selected for this application. Figures \ref{fig:sub_emp_variog_lat} to \ref{fig:sub_emp_variog_lon3} show the resulting semi-variograms and simulation quantiles for the data from four sub-regions defined by the combination of latitudes at north and south of $35$ degrees and longitudes at west and east of $-1$ degree. Each of these four sub-regions contains the same number of data points. The simulation quantiles were constructed  using the method described in \cite{myllymaki2017global}. Figure \ref{fig:sub_emp_variog_iso} suggests a rather poor fit, whilst Figures \ref{fig:sub_emp_variog_lat} to \ref{fig:sub_emp_variog_lon3} collectively indicate non-isotropic behaviour, in that the model fit is tolerable in the first quadrant,  better in the second and third quadrants, but very poor in the fourth quadrant. Similar results obtained for the circular and chordal Mat{\'e}rn models are available at \texttt{https://github.com/FcoCuevas87/F${}\_{}$Family${}\_{}$cova}. 

There might be more sophisticated strategies to handle this problem: for instance,  building a ${\cal F}$ class with spatially adaptive parameters to handle the differences for every subregion. In this work, we keep isotropy as in  \cite{jeong-jun2}, and now focus on the predictive performance of the three models listed above.

\begin{table}
\caption{Average ML estimates and log-Likelihood value attained at the optimum. Standard errors are specified in parentheses.}\label{mle}
\centering
\begin{tabular}{ccccc}  \hline \hline 
 Model &       $\widehat{\sigma}_0$  & $\widehat{\alpha}$  &   $\widehat{\nu}$  &   Log-Likelihood    \\ \hline
$\mathcal{F}$-Family &     $1.088$  &   $0.398$  &  $0.721$   &  $-180.963$  \\   
           & (0.058)  &  (0.154) & (0.228)  &  (22.837) \\ \hline         
Circular-Mat\'ern           &    $1.110$   & $0.289$  &  $0.665$   & $-180.761$   \\   
       &   (0.050)&   (0.088)&   (0.193)&   (22.905) \\    \hline       
Mat\'ern  Chordal     &    $1.060$   & $0.291$  &  $0.669$   & $-180.400$ \\    
       & (0.047) &  (0.091) &  (0.192) &  (22.896)  \\    \hline
\hline 
\end{tabular}
\end{table}

 In this regard, for each sample $i=1,\ldots,500,$ we calculate
\begin{equation*}
 \label{rmse}
 \text{RMSE}_{i} = \frac{1}{100}\sum_{j=1}^{100}\bigg( \frac{1}{20} \sum_{k=1}^{20}(Z(\bm{x}_{k}^{(ij)}) - \widehat{Z}(\bm{x}_{k}^{(ij)}))^2 \bigg)^{1/2},
 \end{equation*}
where $\bm{x}^{(ij)}_{k}$ is the location $k$ of the $j$-sample of the validation set for the repetition $i$, and $Z(\bm{x})$ and $\widehat{Z}(\bm{x})$ denote the true and predicted values at the site $\bm{x}$, respectively.   

\begin{figure}
\centering
\includegraphics[height = 7cm, width = 7cm,keepaspectratio]{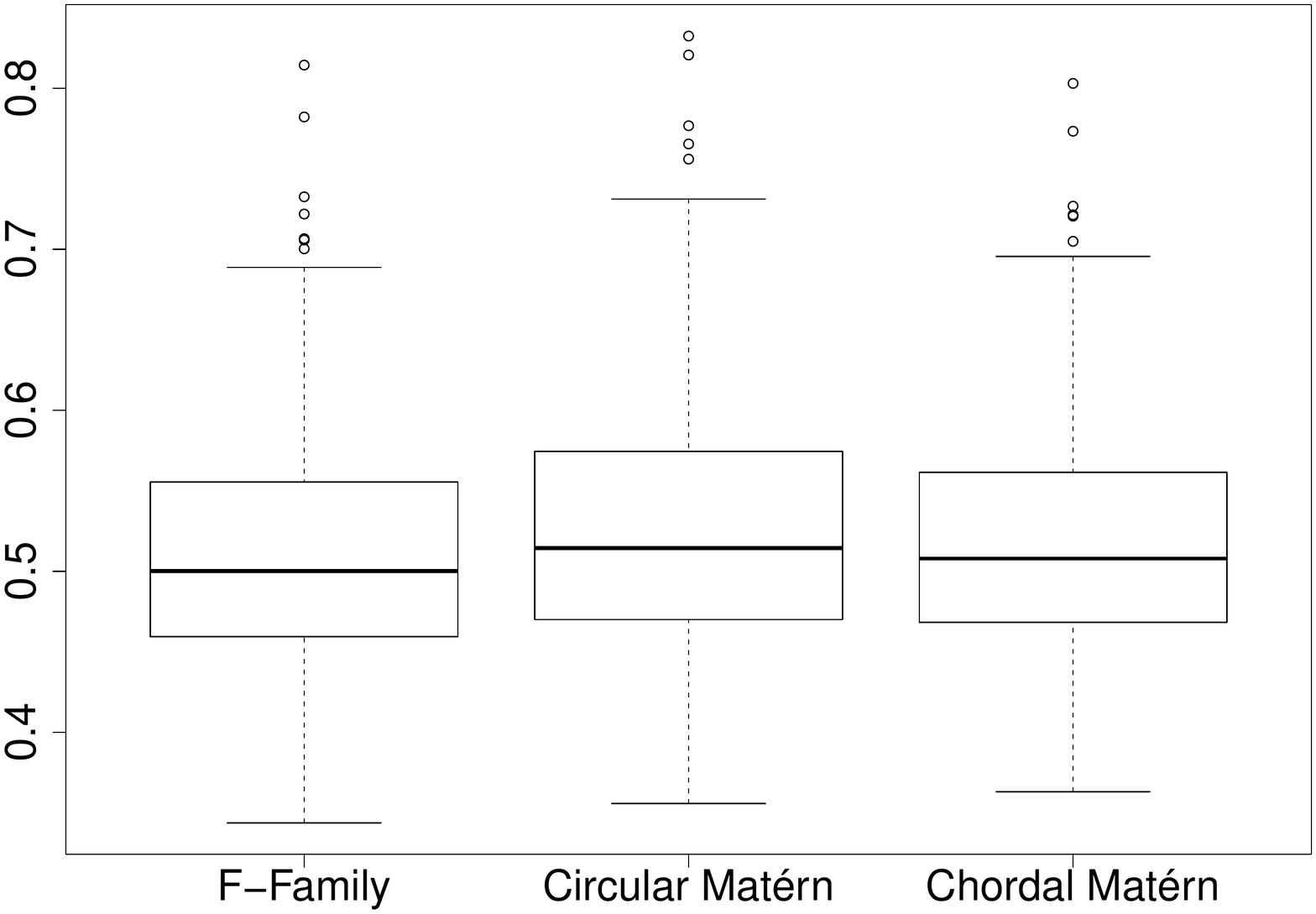} 
\caption{Boxplot of the RMSE for each covariance model.}
\label{pred_boxplots}
\end{figure}

Figure \ref{pred_boxplots} illustrates, for each model, the distribution of RMSE across the $500$ repetitions of the experiment,
whilst Table \ref{prediction} gives their average values. We experience that in $440$ occasions, the $\mathcal{F}$-family of covariance functions gives the lowest values. The average improvements are $2.98\%$ and  $1.39\%$ in terms of relative RMSE, compared with the circular-Mat\'ern and the chordal Mat\'ern models, respectively. 

\begin{table}
\centering
\caption{RMSE averages for each covariance model based on 500 repetitions.}
\label{prediction}
\begin{tabular}{cccccc}  \hline \hline 
                      &  $\mathcal{F}$-Family   &   Circular-Mat\'ern  &  Mat\'ern chordal   \\ \hline
$\text{RMSE}$         &   $\bm{0.510}$    & $0.525$       &   $0.517$    \\  
					  & (0.069)           & (0.074)       &   (0.066)\\                      
\hline
\end{tabular}
\end{table}


\section{Discussion}
\label{sec6}

Our simulation experiment supports the view that the consistency properties of ML estimation under our proposed model mirror known results for the analagous parameterisation of the planar Mat{\'e}rn model. Proving this is a challenging problem. A possible approach would be to use the theory developed by \cite{arafat} about equivalence of Gaussian measures on spherical spaces. This would amount to using the $d$-Schoenberg sequences associated to the ${\cal F}$ class, which are provided in Appendix \ref{A.1}.

The reported differences in predictive performance for the PWC dataset are similar to those reported in other comparative studies \citep[see, {\em e.g.},][]{gneiting2010matern, jeong17}. 

\cite{ste07} proposes  to control the latitude-dependent  variance and the measurement error by augmenting the correlation function with an additive nugget effect. In our experiments, we always found that our estimates for the nugget where identically equal to zero, so that we excluded this from the presentation in Section \ref{sec5}.

Often, climate data are not isotropic on the sphere. In particular, \cite{ste07} evokes \cite{jones} to  call those covariance functions defined over $\S^2$ that are nonstationary over latitudes but stationary over longitudes {\it axially symmetric}. Appendix  \ref{A.2} shows that the ${\cal F}$-Family introduced in this paper can be used as a building block to create models that satisfy axial symmetry.

\section*{Acknowledgement}

Alfredo Alegr\'ia is partially supported by the Commission for Scientific and Technological Research CONICYT, through grant CONICYT/FONDECYT/INICIACi\'ON/No. 11190686. Francisco Cuevas-Pacheco has been supported by the AC3E, UTFSM, under grant FB-0008

\section*{Appendices}
\addcontentsline{toc}{section}{Appendices}
\renewcommand{\thesubsection}{\Alph{subsection}}
\makeatletter\@addtoreset{equation}{subsection}
\def\theequation{\thesubsection.\arabic{equation}}

\subsection{$d$-Schoenberg coefficients for the ${\cal F}$-family} \label{A.1}

Let $p$ and $q$ be positive integers. 
The generalised hypergeometric functions \citep[][15.1.1]{abramowitz1964handbook} are defined as
\begin{equation*}
\pFq{p}{q}{a_{1},\cdots,a_{p}}{b_{1},\cdots,b_{q}}{z} = \sum_{n=0}^{\infty} \frac{(a_{1})_{n} \cdots (a_{p})_{n}}{(b_{1})_{n} \cdots (b_{q})_{n}} \frac{z^{n}}{n!}, \qquad |z| < 1. \\
\end{equation*}
The special case $\pFq{2}{1}{a,b}{c}{z}=  {}_{2}F_{1}(a,b,c;z)$ has been used in Section \ref{sec3} to introduce the ${\cal F}$-Family in Equation (\ref{cov_2f1}).

Let $d $ be a positive integer. We now provide detailed calculations for the $d$-Schoenberg coefficients associated to (\ref{cov_2f1})  through the identity (\ref{d-schoenberg}). Theorem \ref{teo1} has provided an expression for the Schoenberg coefficients related to the Hilbert sphere $\S^{\infty}$. 
\vspace{0.2cm}

\noindent {\bf Property A.1}
Let ${\cal F}_{\tau,\alpha,\nu}$ be the family of functions defined through Equation (\ref{cov_2f1}). Let $d$ be a positive integer. Then, the $d$-Schoenberg sequence of coefficients $\{b_{n,d} \}_{n=0}^{\infty}$ related to ${\cal F}_{\tau,\alpha,\nu}$ through Equation (\ref{d-schoenberg}) are uniquely determined by
$$b_{n,d} = \left\{\begin{array}{cc}
K_{even}(\alpha,\tau,\nu,n,d) \pFq{5}{4}{\frac{\alpha}{2} + n, \frac{\alpha+1}{2} + n, \frac{\tau}{2} + n, \frac{\tau + 1}{2} + n, 1}{\frac{\alpha + \nu + \tau}{2} + n, \frac{\alpha + \nu + \tau+1}{2} + n, \frac{n}{2} + 1, \frac{3n + d + 1}{2}}{1}  & \mbox{ if n is even},\\
& \\
& \\
K_{odd}(\alpha,\tau,\nu,n,d) \pFq{5}{4}{\frac{\alpha}{2} + n + 1, \frac{\alpha+1}{2} + n, \frac{\tau+1}{2} + n, \frac{\tau}{2} + n + 1, 1}{\frac{\alpha + \nu + \tau + 1}{2} + n, \frac{\alpha + \nu + \tau}{2} + n + 1, \frac{n+3}{2}, \frac{3n + d + 2}{2}}{1} & \mbox{ if n is odd},
\end{array}\right.$$
where
$$K_{even}(\alpha,\tau,\nu,n,d) = \frac{B(\alpha,\nu+\tau)}{B(\alpha,\nu)}\frac{(\alpha)_{2n}(\tau)_{2n} (2n+d-1)\Gamma\left(\frac{d-1}{2}\right)}{(\alpha+\nu+\tau)_{2n} 2^{2n+1}\Gamma\left(\frac{n+2}{2}\right)\Gamma\left(\frac{3n+d+1}{2}\right)} {{n+d-2}\choose{n}}, $$ and
$$K_{odd}(\alpha,\tau,\nu,n,d) = \frac{B(\alpha,\nu+\tau)}{B(\alpha,\nu)}\frac{(\alpha)_{2n+1}(\tau)_{2n+1} (2n+d-1)\Gamma\left(\frac{d-1}{2}\right)}{(\alpha+\nu+\tau)_{2n+1} 2^{2n+2}\Gamma\left(\frac{n+3}{2}\right)\Gamma\left(\frac{3n+d+2}{2}\right)} {{n+d-2}\choose{n}}.$$

\noindent {\bf Proof}. The proof of Theorem \ref{teo1} has shown that the Schoenberg coefficients related to the $\Psi_{\infty}$ representation of the family ${\cal F}$ are defined by
\begin{align}\label{schoemberg_seq}
b_{n} = \frac{B(\alpha,\nu+\tau)}{B(\alpha,\nu)}\frac{(\alpha)_{n}(\tau)_{n}}{(\alpha+\nu+\tau)_{n} (n!)}.
\end{align}
Theorem 4.1 in \cite{moller} shows that, since the function ${\cal F}_{\tau, \alpha,\nu}$ belongs to the class $\Psi_{\infty}$, its $d$-Schoemberg coefficients $b_{n,d}$ are uniquely determined through
\begin{align}\label{d_schoemberg}
b_{n,d} = \sum_{\substack{l=n\\ n-l \equiv 0(\mod 2)}}^{\infty} b_{n}\gamma_{n,l}^{(d)}
\end{align}
 where $b_{n}$ is defined in (\ref{schoemberg_seq}) and 
$$\gamma_{n,l}^{d} = \frac{(2n + d - 1)(l!) \Gamma\left(\frac{d-1}{2}\right)} {{2^{l+1} \left\{\left(\frac{l-n}{2}\right)!\right\} \Gamma\left(\frac{l+n+d+1}{2}\right)}} {{n+d-2}\choose{n}}.$$
We can now plugin \eqref{schoemberg_seq} into \eqref{d_schoemberg} to get
\begin{align*}
b_{n,d} &= K_{1}(\alpha,\tau,\nu,n,d) \sum_{\substack{l=n\\ n-l \equiv 0(\mod 2)}} \frac{(\alpha)_{l}(\tau)_{l}}{(\alpha+\nu+\tau)_{l} 2^{l+1} \left\{\left(\frac{l-n}{2}\right)!\right\} \Gamma\left(\frac{l+n+d+1}{2}\right) } \noindent \\
&= K_{1}(\alpha,\tau,\nu,n,d) S_{n}(\alpha,\tau,\nu,n,d),
\end{align*}
 where $$K_{1}(\alpha,\tau,\nu,n,d)= (2n + d - 1)\Gamma\left(\frac{d-1}{2}\right) {{n+d-2}\choose{n}} \frac{B(\alpha,\nu+\tau)}{B(\alpha,\nu)}.$$ When $n$ is an even positive integer, we get that
\begin{align*}
S_{n}(\alpha,\tau,\nu,n,d) = \sum_{j=n}^{\infty} \frac{(\alpha)_{2j}(\tau)_{2j}}{(\alpha+\nu+\tau)_{2j} 2^{2j+1} \left\{\left(\frac{2j-n}{2}\right)!\right\} \Gamma\left(\frac{2j+n+d+1}{2}\right) }.
\end{align*}
We now define $i = j-n$ to obtain 
\begin{align*}
S_{n}(\alpha,\tau,\nu,n,d) = \sum_{i=0}^{\infty} \frac{(\alpha)_{2i + 2n}(\tau)_{2i + 2n}}{(\alpha+\nu+\tau)_{2i + 2n} 2^{2i+2n+1} \left\{\left(i +\frac{n}{2}\right)!\right\} \Gamma\left( i + \frac{3n+d+1}{2}\right) }.
\end{align*}
We can now make use of the factorisation $(\alpha)_{2i+2n} = (\alpha)_{2n}(a+n)_{2i}$ \citep{prudnikov} 
 to obtain
\begin{align*}
S_{n}(\alpha,\tau,\nu,n,d) = \frac{(\alpha)_{2n}(\tau)_{2n}}{(\alpha+\nu+\tau)_{2n}2^{2n+1}}\sum_{i=0}^{\infty} \frac{(\alpha+n)_{2i}(\tau+n)_{2i}}{(\alpha+\nu+\tau+n)_{2i} 2^{2i+1} \left\{\left(i+\frac{n}{2}\right)!\right\} \Gamma\left(i+\frac{3n+d+1}{2}\right) }.
\end{align*}
Using the dimidiation formula for the Pochhammer symbol \citep{prudnikov} 
$$(\alpha)_{2i} = 2^{2i}(\alpha/2)_{i}((\alpha+1)/2)_{i},$$ 
and completing terms, the series $S_{n}$ is
\begin{align*}
S_{n}(\alpha,\tau,\nu,n,d) &= K_{2}(\alpha,\tau,\nu,n,d) \sum_{i=0}^{\infty} \frac{(\frac{\alpha+1}{2}+n)_{i}(\frac{\alpha}{2}+n)_{i}(\frac{\tau}{2}+n)_{i}(\frac{\tau+1}{2}+n)_{i}(1)_{i}}{\left(\frac{\alpha+\nu+\tau}{2}+n\right)_{i}(\frac{\alpha+\nu+\tau+1}{2}+n)_{i}\left(\frac{n}{2}+1\right)_{i}\left(\frac{3n+d+1}{2}\right)_{i} i! },\\
&= K_{2}(\alpha,\tau,\nu,n,d)\pFq{5}{4}{\frac{\alpha}{2} + n, \frac{\alpha+1}{2} + n, \frac{\tau}{2} + n, \frac{\tau + 1}{2} + n, 1}{\frac{\alpha + \nu + \tau}{2} + n, \frac{\alpha + \nu + \tau+1}{2} + n, \frac{n}{2} + 1, \frac{3n + d + 1}{2}}{1},
\end{align*}
where $$ K_{2}(\alpha,\tau,\nu,n,d)  = \frac{\left(\alpha \right)_{2n}(\tau)_{2n}}{(\alpha+\nu+\tau)_{2n}2^{2n+1}\Gamma\left(\frac{n}{2}\right)\Gamma\left(\frac{3n+d+1}{2}\right)},$$ and 
$$K_{even}(\alpha,\tau,\nu,n,d) = K_{1}(\alpha,\tau,\nu,n,d) K_{2}(\alpha,\tau,\nu,n,d).$$ 
When $n$ is an odd positive integer, the proof works \textit{mutatis mutandis} through similar calculations \hfill $\Box$ \\ 


\subsection{Axially symmetric version of the ${\cal F}$ class} \label{A.2}

For phenomena covering a big portion of our planet, isotropy is a questionable assumption.  On the one hand, isotropy might be expected for microscale metereology on a sufficiently temporally aggregated level for many physical quantities. On the other hand, mesoscale and synoptic scale meteorology are not even approximately isotropic, due to the highly nonlinear nature of the Earth's system. Indeed, \cite{ste07} shows that  total column ozone data show significant changes over latitude. \cite{castruccio1} argued that both inter and intra annual variability for surface temperature is depend on latitude. For the sequel, we refer to the unit sphere $\mathbb{S}^2$ of $\R^3$ with coordinates $\bm{x}=(L,\ell)^{\top}$, with $L \in [0,\pi]$ denoting latitude and $\ell \in [0, 2 \pi)$ denoting longitude. 
In particular, \cite{ste07} resorts to the results in \cite{jones} to call the covariance $C$ axially symmetric when 
\begin{equation*} \label{def:axsym}
 C\left ( \bm{x}_1, \bm{x}_2 \right )  = {\cal C}(L_1,L_2, \ell_1-\ell_2), \qquad (L_i,\ell_i) \in [0,\pi] \times [0,2 \pi) , i=1,2. 
\end{equation*}
Axially symmetric processes have a well understood spectral representation that includes as a special case the geodesic isotropy illustrated through Equations (\ref{isotropy}) and (\ref{spectral_rep}). For details, the reader is referred to \cite{jones} and more recently to \cite{ste07}. 

The literature on axially symmetric models is sparse, with the attempt in \cite{PCAP} being a notable exception. Let $d_{{\rm CH}}(\ell_1,\ell_2)$ denote the chordal distance between two longitudes $\ell_1$ and $\ell_2$. Let ${\cal M}_{\alpha,\nu}$ denote the Mat{\'e}rn class defined at (\ref{matern}). Then, \cite{PCAP} propose an axially symmetric model of the type
\begin{equation*}
\label{matern_ax_sym}
{\cal C}(L_1,L_2, \ell_1-\ell_2) = \sigma(L_1,L_2) {\cal M}_{\alpha(L_1,L_2), \nu(L_1,L_2)} \left ( d_{{\rm CH}}(\ell_1,\ell_2  )\right ), 
\end{equation*}
$(L_i,\ell_i) \in [0,\pi] \times [0,2 \pi) , i=1,2$, where $\sigma, \alpha$ and $\nu$ are strictly positive functions that must be carefully chosen in order to preserve positive definiteness. The interpretation of these functions is very intuitive, as they indicate how, respectively, variance, scale and smoothness can vary across latitudes. Usually $\sigma$ is modeled through a linear combination of Legendre polynomials \citep{jun-stein}. To illustrate the new model, we need to define a stochastic process $\{X(L), \; L \in [0,\pi] \}$ and we call variogram the quantity ${\rm Var} \left ( X(L_2) -X(L_1) \right )/2$, $L_1,L_2 \in [0,\pi] $ \citep[see][with the references therein]{chiles}.  \\

\noindent {\bf Theorem B.1} \label{F-axial-symmetry} 
Let ${\cal F}$ be the family of functions defined at Equation (\ref{cov_2f1}). Let $\tau>0$.  Let $\sigma: [0,\pi]^2 \to \R_+$ be positive definite and let $\alpha, \nu: [0, \pi]^2 \to \R_+ $ be continuous functions such that the functions 
$(L_1,L_2) \mapsto {\alpha(L_1,L_2)}$ and $(L_1,L_2) \mapsto {\nu(L_1,L_2)}$ define two variograms on $[0,\pi]^2$. Then, the function
\begin{equation*}
\label{cov-F-AS} 
{\cal C}(L_1,L_2, \ell_1-\ell_2) = \sigma(L_1,L_2) {\cal F}_{\tau, \alpha(L_1,L_2),\nu(L_1,L_2)} \left ( \theta(\ell_1,\ell_2) \right ),  
\end{equation*}
for $(L_i,\ell_i) \in [0,\pi] \times [0,2 \pi)$, is positive definite.

\noindent {\bf Proof}. We give a constructive proof. We consider the scale mixture
\begin{equation*}
\label{scale2} 
\int_{0}^{1} \frac{(1-\delta)^{\tau}}{\left ( 1 - \delta \cos \theta (\ell_1,\ell_2)  \right )^{\tau}} \delta^{\alpha(L_1,L_2)-1} (1- \delta)^{\nu(L_1,L_2)-1} {\rm d} \delta.
\end{equation*}
Clearly, the function $(\ell_1,\ell_2) \mapsto (1-\delta)^{\tau}/(1- \delta \cos \theta (\ell_1,\ell_2))^{\tau}$ is positive definite for any $\tau >0$ and $\delta \in (0,1)$. Both functions $a^{\alpha}$ and $a^{\nu}$ are positive definite on $[0,\pi]^2$ provided $0 \le a \le 1$ \citep[direct consequence of ][Theorem 2]{schoenberg}. Since the scale mixture above is well defined, the proof is completed by using the same arguments as in Theorem \ref{teo1}. \hfill $\Box$

\bibliographystyle{chicago}

\bibliography{Bibliography_cor}

\begin{thebibliography}{}

\bibitem[\protect\citeauthoryear{Abramowitz and Stegun}{Abramowitz and
  Stegun}{1965}]{abramowitz1964handbook}
Abramowitz, M. and I.~Stegun (1965).
\newblock {\em Handbook of {M}athematical {F}unctions: with {F}ormulas,
  {G}raphs, and {M}athematical {T}ables}, Volume~55.
\newblock Courier Corporation.

\bibitem[\protect\citeauthoryear{Arafat, Porcu, Bevilacqua, and Mateu}{Arafat
  et~al.}{2018}]{arafat}
Arafat, A., E.~Porcu, M.~Bevilacqua, and J.~Mateu (2018).
\newblock Equivalence and orthogonality of {G}aussian measures on spheres.
\newblock {\em Journal of Multivariate Analysis\/}~{\em 267}, 306--318.

\bibitem[\protect\citeauthoryear{Banerjee}{Banerjee}{2005}]{banerjee}
Banerjee, S. (2005).
\newblock On geodetic distance computations in spatial modeling.
\newblock {\em Biometrics\/}~{\em 61}, 617--625.

\bibitem[\protect\citeauthoryear{Beatson, zu~Castell, and Xu}{Beatson
  et~al.}{2014}]{beatson}
Beatson, R.~K., W.~zu~Castell, and Y.~Xu (2014).
\newblock P{\'o}lya criterion for (strict) positive definiteness on the sphere.
\newblock {\em IMA Journal of Numerical Analysis\/}~{\em 34}, 550--568.

\bibitem[\protect\citeauthoryear{Berg and Porcu}{Berg and
  Porcu}{2017}]{berg-porcu}
Berg, C. and E.~Porcu (2017).
\newblock From {S}choenberg coefficients to {S}choenberg functions.
\newblock {\em Constructive Approximation\/}~{\em 45}, 217--241.

\bibitem[\protect\citeauthoryear{Bevilacqua, Gaetan, Mateu, and
  Porcu}{Bevilacqua et~al.}{2012}]{bevb:12}
Bevilacqua, M., C.~Gaetan, J.~Mateu, and E.~Porcu (2012).
\newblock Estimating space and space-time covariance functions: a weighted
  composite likelihood approach.
\newblock {\em Journal of the American Statistical Association\/}~{\em 107},
  268--280.

\bibitem[\protect\citeauthoryear{Castruccio and Stein}{Castruccio and
  Stein}{2013}]{castruccio1}
Castruccio, S. and M.~L. Stein (2013).
\newblock Global space-time models for climate ensembles.
\newblock {\em Annals of Applied Statistics\/}~{\em 7}, 1593--1611.

\bibitem[\protect\citeauthoryear{Chiles and Delfiner}{Chiles and
  Delfiner}{1999}]{chiles}
Chiles, J. and P.~Delfiner (1999).
\newblock {\em Geostatistics: {M}odeling {S}patial {U}ncertainty}.
\newblock New York: Wiley.

\bibitem[\protect\citeauthoryear{Daley and Porcu}{Daley and
  Porcu}{2013}]{daley-porcu}
Daley, D.~J. and E.~Porcu (2013).
\newblock Dimension walks and {S}choenberg spectral measures.
\newblock {\em Proceedings of the American Mathematical Society\/}~{\em 141},
  1813--1824.

\bibitem[\protect\citeauthoryear{Furrer, Genton, and Nychka}{Furrer
  et~al.}{2006}]{FGN}
Furrer, R., M.~G. Genton, and D.~Nychka (2006).
\newblock Covariance tapering for interpolation of large spatial datasets.
\newblock {\em Journal of Computational and Graphical Statistics\/}~{\em 15},
  502--523.

\bibitem[\protect\citeauthoryear{Galassi, Davies, Theiler, Gough, Jungman,
  Alken, Booth, and Rossi}{Galassi et~al.}{1996}]{galassi1996gnu}
Galassi, M., J.~Davies, J.~Theiler, B.~Gough, G.~Jungman, P.~Alken, M.~Booth,
  and F.~Rossi (1996).
\newblock {GNU} scientific library reference manual.

\bibitem[\protect\citeauthoryear{Gneiting}{Gneiting}{2013}]{gneiting2013}
Gneiting, T. (2013).
\newblock Strictly and non-strictly positive definite functions on spheres.
\newblock {\em Bernoulli\/}~{\em 19}, 1327--1349.

\bibitem[\protect\citeauthoryear{Gneiting, Kleiber, and Schlather}{Gneiting
  et~al.}{2010}]{gneiting2010matern}
Gneiting, T., W.~Kleiber, and M.~Schlather (2010).
\newblock Mat{\'e}rn cross-covariance functions for multivariate random fields.
\newblock {\em Journal of the American Statistical Association\/}~{\em
  105\/}(491), 1167--1177.

\bibitem[\protect\citeauthoryear{Guinness and Fuentes}{Guinness and
  Fuentes}{2016}]{guinness}
Guinness, J. and M.~Fuentes (2016).
\newblock Isotropic covariance functions on spheres: some properties and
  modeling considerations.
\newblock {\em Journal of Multivariate Analysis\/}~{\em 143}, 143--152.

\bibitem[\protect\citeauthoryear{Hansen, Thorarinsdottir, Ovcharov, Gneiting,
  and Richards}{Hansen et~al.}{2015}]{hansen2015}
Hansen, L.~V., T.~L. Thorarinsdottir, E.~Ovcharov, T.~Gneiting, and D.~Richards
  (2015).
\newblock Gaussian random particles with flexible {H}ausdorff dimension.
\newblock {\em Advances in Applied Probability\/}~{\em 47\/}(2), 307--327.

\bibitem[\protect\citeauthoryear{Jeong, Castruccio, Crippa, Genton,
  et~al.}{Jeong et~al.}{2018}]{jeong17}
Jeong, J., S.~Castruccio, P.~Crippa, M.~G. Genton, et~al. (2018).
\newblock Reducing storage of global wind ensembles with stochastic generators.
\newblock {\em The Annals of Applied Statistics\/}~{\em 12\/}(1), 490--509.

\bibitem[\protect\citeauthoryear{Jeong and Jun}{Jeong and
  Jun}{2015a}]{jeong-jun}
Jeong, J. and M.~Jun (2015a).
\newblock A class of {M}at{\'e}rn-like covariance functions for smooth
  processes on a sphere.
\newblock {\em Spatial Statistics\/}~{\em 11}, 1--18.

\bibitem[\protect\citeauthoryear{Jeong and Jun}{Jeong and
  Jun}{2015b}]{jeong-jun2}
Jeong, J. and M.~Jun (2015b).
\newblock Covariance models on the surface of a sphere: when does it matter?
\newblock {\em STAT\/}~{\em 4}, 167--182.

\bibitem[\protect\citeauthoryear{Johansson}{Johansson}{2017}]{Johansson2017arb}
Johansson, F. (2017).
\newblock Arb: efficient arbitrary-precision midpoint-radius interval
  arithmetic.
\newblock {\em IEEE Transactions on Computers\/}, 1281--1292.

\bibitem[\protect\citeauthoryear{Johnson, Kemp, and Kotz}{Johnson
  et~al.}{2005}]{johnson2005univariate}
Johnson, N.~L., A.~W. Kemp, and S.~Kotz (2005).
\newblock {\em Univariate {D}iscrete {D}istributions}.
\newblock John Wiley \& Sons, New York.

\bibitem[\protect\citeauthoryear{Johnson, Kotz, and Balakrishnan}{Johnson
  et~al.}{1995}]{johnson1995continuous}
Johnson, N.~L., S.~Kotz, and N.~Balakrishnan (1995).
\newblock {\em Continuous {U}nivariate {D}istributions, vol. 2}.
\newblock Wiley, New York,.

\bibitem[\protect\citeauthoryear{Jones}{Jones}{1963}]{jones}
Jones, R.~H. (1963).
\newblock Stochastic processes on a sphere.
\newblock {\em Annals of Mathematical Statistics\/}~{\em 34}, 213--218.

\bibitem[\protect\citeauthoryear{Jun and Stein}{Jun and
  Stein}{2007}]{jun-stein}
Jun, M. and M.~L. Stein (2007).
\newblock An approach to producing space-time covariance functions on spheres.
\newblock {\em Technometrics\/}~{\em 49}, 468--479.

\bibitem[\protect\citeauthoryear{Kalnay, Kanamitsu, Kistler, Collins, Deaven,
  Gandin, Iredell, Saha, White, Woollen, et~al.}{Kalnay
  et~al.}{1996}]{kalnay1996ncep}
Kalnay, E., M.~Kanamitsu, R.~Kistler, W.~Collins, D.~Deaven, L.~Gandin,
  M.~Iredell, S.~Saha, G.~White, J.~Woollen, et~al. (1996).
\newblock The {NCEP/NCAR} 40-year reanalysis project.
\newblock {\em Bulletin of the American meteorological Society\/}~{\em
  77\/}(3), 437--472.

\bibitem[\protect\citeauthoryear{Kaufman and Shaby}{Kaufman and
  Shaby}{2013}]{kaufman}
Kaufman, C. and B.~Shaby (2013).
\newblock The role of the range parameter for estimation and prediction in
  geostatistics.
\newblock {\em Biometrika\/}~{\em 100}, 473--484.

\bibitem[\protect\citeauthoryear{Lang and Schwab}{Lang and
  Schwab}{2013}]{lang-schwab}
Lang, A. and C.~Schwab (2013).
\newblock Isotropic random fields on the sphere: regularity, fast simulation
  and stochastic partial differential equations.
\newblock {\em Annals of Applied Probabilty\/}~{\em 25}, 3047--3094.

\bibitem[\protect\citeauthoryear{Lin, Mu, Cheung, Dunson, et~al.}{Lin
  et~al.}{2019}]{lin2019extrinsic}
Lin, L., N.~Mu, P.~Cheung, D.~Dunson, et~al. (2019).
\newblock Extrinsic gaussian processes for regression and classification on
  manifolds.
\newblock {\em Bayesian Analysis\/}~{\em 14\/}(3), 907--926.

\bibitem[\protect\citeauthoryear{Lindgren, Rue, and Lindstroem}{Lindgren
  et~al.}{2011}]{Lindgren}
Lindgren, F., H.~Rue, and J.~Lindstroem (2011).
\newblock An explicit link between {G}aussian fields and {G}aussian {M}arkov
  random fields: the stochastic partial differential equation approach.
\newblock {\em Journal of the Royal Statistical Society: Series B\/}~{\em 73},
  423--498.

\bibitem[\protect\citeauthoryear{Massa, Per{\'o}n, and Porcu}{Massa
  et~al.}{2017}]{massa}
Massa, E., A.~Per{\'o}n, and E.~Porcu (2017).
\newblock Positive definite functions on complex spheres, and their walks
  through dimensions.
\newblock {\em SIGMA\/}~{\em 13}.

\bibitem[\protect\citeauthoryear{Menegatto, Oliveira, and Per{\'o}n}{Menegatto
  et~al.}{2006}]{menegatto-peron}
Menegatto, V.~A., C.~P. Oliveira, and A.~P. Per{\'o}n (2006).
\newblock Strictly positive definite kernels on subsets of the complex plane.
\newblock {\em Computional Mathematics and Applications\/}~{\em 51},
  1233--1250.

\bibitem[\protect\citeauthoryear{M{\o}ller, Nielsen, Porcu, and
  Rubak}{M{\o}ller et~al.}{2018}]{moller}
M{\o}ller, J., M.~Nielsen, E.~Porcu, and E.~Rubak (2018).
\newblock Determinantal point process models on the sphere.
\newblock {\em Bernoulli\/}~{\em 24\/}(2), 1171--1201.

\bibitem[\protect\citeauthoryear{M{\"u}ller and Thompson}{M{\"u}ller and
  Thompson}{2016}]{muller2016comparing}
M{\"u}ller, M. and S.~Thompson (2016).
\newblock Comparing statistical and process-based flow duration curve models in
  ungauged basins and changing rain regimes.
\newblock {\em Hydrology and Earth System Sciences\/}~{\em 20\/}(2), 669.

\bibitem[\protect\citeauthoryear{M{\"u}ller, Dralle, and Thompson}{M{\"u}ller
  et~al.}{2014}]{muller2014analytical}
M{\"u}ller, M.~F., D.~N. Dralle, and S.~E. Thompson (2014).
\newblock Analytical model for flow duration curves in seasonally dry climates.
\newblock {\em Water Resources Research\/}~{\em 50\/}(7), 5510--5531.

\bibitem[\protect\citeauthoryear{Myllym{\"a}ki, Mrkvi{\v{c}}ka, Grabarnik,
  Seijo, and Hahn}{Myllym{\"a}ki et~al.}{2017}]{myllymaki2017global}
Myllym{\"a}ki, M., T.~Mrkvi{\v{c}}ka, P.~Grabarnik, H.~Seijo, and U.~Hahn
  (2017).
\newblock Global envelope tests for spatial processes.
\newblock {\em Journal of the Royal Statistical Society: Series B (Statistical
  Methodology)\/}~{\em 79\/}(2), 381--404.

\bibitem[\protect\citeauthoryear{Olver, Lozier, Boisvert, and Clark}{Olver
  et~al.}{2010}]{olver2010nist}
Olver, F.~W., D.~W. Lozier, R.~F. Boisvert, and C.~W. Clark (2010).
\newblock {\em {NIST} handbook of mathematical functions hardback and
  {CD-ROM}}.
\newblock Cambridge university press, New York.

\bibitem[\protect\citeauthoryear{Porcu, Alegr{\'i}a, and Furrer}{Porcu
  et~al.}{2018}]{PAF2016}
Porcu, E., A.~Alegr{\'i}a, and R.~Furrer (2018).
\newblock Modeling temporally evolving and spatially globally dependent data.
\newblock {\em International Statistical Review\/}~{\em 86\/}(2), 344--377.

\bibitem[\protect\citeauthoryear{Porcu, Bevilacqua, and Genton}{Porcu
  et~al.}{2016}]{PBG16}
Porcu, E., M.~Bevilacqua, and M.~G. Genton (2016).
\newblock Spatio-temporal covariance and cross-covariance functions of the
  great circle distance on a sphere.
\newblock {\em Journal of the American Statistical Association\/}~{\em
  111\/}(514), 888--898.

\bibitem[\protect\citeauthoryear{Porcu, Castruccio, Alegr{\'\i}a, and
  Crippa}{Porcu et~al.}{2019}]{PCAP}
Porcu, E., S.~Castruccio, A.~Alegr{\'\i}a, and P.~Crippa (2019).
\newblock Axially symmetric models for global data: A journey between
  geostatistics and stochastic generators.
\newblock {\em Environmetrics\/}~{\em 30\/}(1), e2555.

\bibitem[\protect\citeauthoryear{Prudnikov, Brichkov, and Marichev}{Prudnikov
  et~al.}{1983}]{prudnikov}
Prudnikov, A., Y.~Brichkov, and O.~Marichev (1983).
\newblock {\em {I}ntegrals and {S}eries. {S}pecial {F}unctions}.
\newblock Gordon and Breach, New York.

\bibitem[\protect\citeauthoryear{Scheuerer, Schlather, and Schaback}{Scheuerer
  et~al.}{2013}]{SSS}
Scheuerer, M., M.~Schlather, and R.~Schaback (2013).
\newblock Interpolation of spatial data - a stochastic or a deterministic
  problem?
\newblock {\em European Journal of Applied Mathematics\/}~{\em 24}, 601--609.

\bibitem[\protect\citeauthoryear{Schoenberg}{Schoenberg}{1942}]{schoenberg}
Schoenberg, I.~J. (1942).
\newblock Positive definite functions on spheres.
\newblock {\em Duke Math. Journal\/}~{\em 9}, 96--108.

\bibitem[\protect\citeauthoryear{Skorokhod and Yadrenko}{Skorokhod and
  Yadrenko}{1973}]{SY73}
Skorokhod, A.~V. and M.~I. Yadrenko (1973).
\newblock On absolute continuity of measures corresponding to homogeneous
  {G}aussian fields.
\newblock {\em Theory of Probability and Its Applications\/}~{\em 18}, 27--40.

\bibitem[\protect\citeauthoryear{Soubeyrand, Enjalbert, and Sache}{Soubeyrand
  et~al.}{2008}]{soub}
Soubeyrand, S., J.~Enjalbert, and I.~Sache (2008).
\newblock Accounting for roughness of circular processes: using {G}aussian
  random processes to model the anisotropic spread of airborne plant disease.
\newblock {\em Theoretical Population Biology\/}~{\em 73}, 92--103.

\bibitem[\protect\citeauthoryear{Stein}{Stein}{1999}]{stein-book}
Stein, M.~L. (1999).
\newblock {\em Statistical {I}nterpolation of {S}patial {D}ata: {S}ome {T}heory
  for {K}riging}.
\newblock Springer, New York.

\bibitem[\protect\citeauthoryear{Stein}{Stein}{2007}]{ste07}
Stein, M.~L. (2007).
\newblock Spatial variation of total column ozone on a global scale.
\newblock {\em Annals of Applied Statistics\/}~{\em 1}, 191--210.

\bibitem[\protect\citeauthoryear{Verbyla}{Verbyla}{1993}]{verbyla1993modelling}
Verbyla, A.~P. (1993).
\newblock Modelling variance heterogeneity: residual maximum likelihood and
  diagnostics.
\newblock {\em Journal of the Royal Statistical Society: Series B
  (Methodological)\/}~{\em 55\/}(2), 493--508.

\bibitem[\protect\citeauthoryear{White and Porcu}{White and Porcu}{2018}]{WP}
White, P. and E.~Porcu (2018).
\newblock Towards a complete picture of stationary covariance functions on
  spheres cross time.
\newblock {\em arXiv preprint arXiv:1807.04272\/}.

\bibitem[\protect\citeauthoryear{Whittaker and Watson}{Whittaker and
  Watson}{1996}]{whittaker1996course}
Whittaker, E.~T. and G.~N. Watson (1996).
\newblock {\em A {C}ourse of {M}odern {A}nalysis}.
\newblock Cambridge university press, Cambridge.

\bibitem[\protect\citeauthoryear{Zhang}{Zhang}{2004}]{zhang}
Zhang, H. (2004).
\newblock Inconsistent estimation and asymptotically equal interpolations in
  model-based geostatistics.
\newblock {\em Journal of the American Statistical Association\/}~{\em 99},
  250--261.

\bibitem[\protect\citeauthoryear{Ziegel}{Ziegel}{2014}]{ziegel}
Ziegel, J. (2014).
\newblock Convolution roots and differentiability of isotropic positive
  definite functions on spheres.
\newblock {\em Proceedings of the American Mathematical Society\/}~{\em 142},
  2053--2077.

\end{thebibliography}

\end{document}